\documentclass[a4paper,fleqn,usenatbib]{mnras}

\usepackage[T1]{fontenc}
\usepackage{ae,aecompl}

\usepackage{graphicx}	
\usepackage{amsmath}	
\usepackage{amssymb}	
\usepackage{float}
\usepackage{array}
\usepackage{placeins}
\usepackage[table]{xcolor}
\usepackage{multirow}
\usepackage{bigstrut}
\usepackage{siunitx}

\graphicspath{{./}}

\def\xmm{\textit{XMM-Newton} }
\def\astrosat{\textit{AstroSat} }
\def\swift{\textit{Swift} }
\makeatletter
    \def\fps@figure{h!tbp}
    \def\fps@table{h!tbp}
\makeatother
\DeclareSIUnit\angstrom{\text {Å}}

\title[Timescale-dependent lags of NGC 4593]{Timescale-dependent X-ray to UV time lags of NGC 4593 using high-intensity \textit{XMM-Newton} observations with \textit{Swift} and \textit{AstroSat}}

\author[Beard, M. W. J. et al.]{
Max W. J. Beard,$^{1}$\thanks{E-mail: mb15g14@soton.ac.uk (MWJB)}
Ian M. M$\rm^{c}$Hardy,$^{1}$
Kavita Kumari,$^{2}$
Gulab C. Dewangan,$^{2}$ Iossif\newauthor
\hspace{4pt}Papadakis,$^{3}$
Dipankar Bhattacharya,$^{2}$
Kulinder Pal Singh,$^{4}$
Daniel Kynoch,$^{1}$
Mayukh Pahari$^{1}$
\\
\\
$^{1}$University of Southampton, University Road, Southampton SO17 1BJ, UK\\
$^{2}$IUCAA, Pune University Campus, Ganeshkhind, Pune 411007, Maharashtra, India\\
$^{3}$University of Crete, Voutes University Campus, GR-70013 Heraklion, Greece\\
$^{4}$IISER Mohali, Knowledge City, Sector 81, Manauli PO, SAS Nagar, Punjab 140306, India\\ 
}

\date{Accepted XXX. Received YYY; in original form ZZZ}

\pubyear{2022}

\begin{document}
\label{firstpage}
\pagerange{\pageref{firstpage}--\pageref{lastpage}}
\maketitle

\begin{abstract}
We present a 140ks observation of NGC 4593 with \xmm providing simultaneous and continuous PN X-ray and OM UV (UVW1 2910\AA) lightcurves which sample short-timescale variations better than previous observations. These observations were simultaneous with 22d of \swift X-ray and UV/optical monitoring, reported previously, and 4d of \astrosat X-ray (SXT), far (FUV 1541\AA), and near (NUV 2632\AA) UV allowing lag measurements between them and the highly-sampled XMM. From the XMM we find that UVW1 lags behind the X-rays by 29.5$\pm$1.3ks, $\sim$half the lag previously determined from the \swift monitoring. Re-examination of the \swift data reveals a bimodal lag distribution, with evidence for both the long and short lags. However if we detrend the \swift lightcurves by LOWESS filtering with a 5d width, only the shorter lag (23.8$\pm$21.2ks) remains. The NUV observations, compared to PN and SXT, confirm the $\sim30$ks lag found by XMM and, after 4d filtering is applied to remove the long-timescale component, the FUV shows a lag of $\sim23$ks. The resultant new UVW1, FUV, and NUV lag spectrum extends to the X-ray band without requiring additional X-ray to UV lag offset, which if the UV arises from reprocessing of X-rays, implies direct illumination of the reprocessor. By referencing previous \swift and HST lag measurements, we obtain an X-ray to optical lag spectrum which agrees with a model using the KYNreverb disc-reprocessing code, assuming the accepted mass of $7.63\times10^{6}M_{\odot}$ and a spin approaching maximum. Previously noted lag contribution from the BLR in the Balmer and Paschen continua are still prominent.

\end{abstract}


\clearpage

\section{Introduction} \label{intro}

Understanding the inner structures of AGN is one of the main aims of extragalactic research. However in almost all cases, the AGN are too small to resolve by direct imaging and so other techniques must be used. One particularly useful technique is 'reverberation mapping' \citep{blandford_mckee_1982} where we measure the time lags between the X-ray emission and that in longer wavelength UV and optical wavebands. The assumption is that the X-rays originate from around the central supermassive black hole (SMBH) and illuminate the surrounding material, principally the accretion disc and broad line region (BLR) gas. These structures reprocess the X-rays and re-emit at longer UV/optical wavelengths. \par 

In the absence of external illumination, the accretion disc will have a radial temperature profile which depends on its mass, spin and accretion rate and is given by \cite{shakura_1973_black} or, with relativistic corrections, by \cite{novikov_thorne_1973}. X-ray illumination increases the temperature and boosts the emission. \par

However unless the X-ray illumination on unit area of the disc is comparable to the intrinsic black body emission from the disc, which almost never happens, the centroid of emission at any particular wavelength only moves inwards by a per cent or two. Thus the X-ray to UV/optical lags can be used to map out the temperature structure of the disc. For reprocessing from the BLR both UV/optical line and continuum emission is produced \citep{korista_goad_2001,korista_goad_2019}, with peaks in the u-band due to the Balmer continuum and in the i-band due to the Paschen continuum. The u and i-band lags produced by the BLR are longer than for those produced by the accretion disc as the BLR lies further from the SMBH. \par 

There have now been many studies of correlated X-ray and UV/optical variability. Early studies combined X-ray observations from RXTE with ground based optical observations \citep[e.g.][]{uttley_2003,suganuma_2006,arevalo_2009,breedt_2009,breedt_2010}. These studies typically showed that the X-rays led the optical by about one day, consistent with reprocessing, but with large uncertainty ($\sim$0.5 d) and without sufficient wavelength detail to map out the temperature structure of the reprocessor. \par 

Later studies, based around X-ray/UV/optical monitoring with \swift provided much greater wavelength coverage \citep[e.g.][]{cameron_2012_correlated,shappee_2014,mchardy_2014_swift,edelson_2015_space,mchardy_et_al_2018,cackett_2018,edelson_2019,vincentelli_2021,vincentelli_2022}, broadly confirming reprocessing from both disc and BLR. However a number of problems with the simple reprocessing scenario emerged. Here we concentrate on just one of these problems, which we shall call 'the timescale problem', i.e. the fact that the lag that is measured depends on the time resolution of the data and whether you have removed long timescale trends or not. This problem is particularly apparent between the X-ray and far-UV bands. \par

It has been noted, since the earliest \swift papers \citep[e.g.][]{shappee_2014}, that although a lag spectrum of the form lag $\propto \lambda^{4/3}$, expected from reprocessing from an accretion disc \citep[e.g.][]{cackett_2007_testing}, was usually a good fit to the lags throughout the UV and optical bands, the extrapolation of that fit back to the X-ray band (i.e. $\sim$2\AA) predicted an X-ray to UV lag which was a good deal smaller than the one that was actually observed (see also summary in \cite{mchardy_et_al_2018}). Thus the lag spectrum is usually steeper between the X-rays and far-UV than between the UV and optical bands.\par

A variety of solutions have been proposed, including that the X-rays are scattered through an inflated inner edge of the disc, thus introducing an additional delay, before emerging as far-UV photons which then illuminate the outer disc \citep{gardner_2017_the}. Alternatively, including a more distant reprocessor than the disc, i.e. the BLR, can produce the additional lag in some cases \citep{cackett_2018, mchardy_et_al_2018}. It has also recently been shown \citep{mchardy_beard_2022} that an accretion disc, truncated at the dust sublimation radius, can explain a steeper X-ray to far-UV lag spectrum.\par

Early on \citep{mchardy_2014_swift} it was noted that if the lightcurves were filtered to remove long timescale trends, the X-ray to UV lags derived from the resultant detrended lightcurves were quite consistent with an extrapolation of the lags within the UV/optical bands back to the X-ray band. Long timescale trends are more apparent in the UV/optical bands than in the X-rays and indicate a second source of slower variations which  affects the UV/optical bands but has less effect in the X-ray band. This second source of variability could be accretion rate variations propagating inwards through the disc \citep[][]{arevalo_2009,breedt_2009}. \par

Further examples of how the X-ray to far-UV lag is reduced by the removal of long term trends are given in the case of NGC~4593, the source discussed in the present paper, by \cite{mchardy_et_al_2018}  and also, for NGC~7469, by \cite{pahari_2020}. The latter case is particularly dramatic. From a month or so of almost continuous X-ray observations with RXTE and UV observations by IUE, \cite{nandra_1998} had shown that, in the original, non-detrended lightcurves, the UV led the X-rays by about 4 days, which is entirely contrary to the expectations of X-ray reprocessing. However after removing only the longer timescales ($>$5 d), \cite{pahari_2020} showed that the far-UV lagged behind the X-rays by 0.37 d and that the longer wavelength UV lagged the short wavelength by similarly short amounts. The detrended lag results were entirely consistent with reprocessing from an accretion disc. Thus the reprocessing signal was present but had been hidden by the second source of variations. \par

\cite{welsh_1999} provides a detailed analysis of the errors that can occur when measuring lags from cross-correlation functions if long timescale trends are not removed. Care, of course, must be taken not to filter out timescales of interest. It is noted, entirely predictably, by \cite{mchardy_et_al_2018} that the measured lags decrease as the timescale of filtering decreases. Thus signals of reprocessing from the BLR, which can be seen in the tails to longer timescales in the MEMecho \citep{horne_2004} UV and optical response functions of NGC~4593 \citep{mchardy_et_al_2018}, may be filtered out with over-vigorous filtering. \par

The analysis described above is important because, if the lags are consistent with simple disc reprocessing without need for an additional X-ray to UV lag, it shows that the accretion disc must be directly visible, without obscuration, to at least part of the X-ray emission region. Thus the inner disc cannot be too highly inflated, and remain optically thick to X-rays. \par

NGC~4593 is one of the brightest Seyfert galaxies in the sky and has been subject to intense monitoring with \swift \citep{mchardy_et_al_2018} and HST \citep{cackett_2018}. Whilst the lags between the various UV and optical bands were more or less in agreement with simple disc reprocessing, the observed lag between the X-rays and the shortest wavelength UV band, i.e. UVW2 (1920 \AA) was $\sim$66 ks compared to the model lag based on reprocessing from the accretion disc of $\sim$8 ks.  This difference causes a big problem for simple disc reprocessing models although it was shown that it was possible to reproduce the observed UV/optical lightcurves from the observed X-ray lightcurve as long as the reprocessor had long tail, probably consistent with additional reprocessing from the BLR \citep{mchardy_et_al_2018}. \par

The range of lags which is accessible depends on the frequency of sampling. \swift observed NGC~4593 almost every orbit (96 min) for 6.4 d from 2016 July 13 to 18 and thereafter every second orbit for a further 16.2 d. Each observation totalled approximately 1 ks although observations were often split into two, or sometimes more, visits. Whilst the sampling rate for the first 6 d is very good by \swift standards, the observations were far from continuous, limiting our coverage of short timescale variations. \par

However, during the 6.4 d of intensive \swift observations, we were able to arrange a continuous 140 ks (1.6 d) observation with \xmm, providing continuous X-ray coverage and also continuous UV coverage (2910\AA), using the Optical Monitor in Fast Mode. These observations provide superb coverage of short timescale variability. We also observed for $\sim$4 of the 6.4 d period with \astrosat, providing additional X-ray coverage and also coverage in two further UV bands. In this paper we re-examine the lags between the X-ray and UV emission in NGC~4593 (Section~\ref{Analysis}) in the light of this additional data and by re-examining the \swift data. We find that the measured lag does indeed depend very significantly on the data and the timescales available for investigation within that data. We conclude by fitting disc reprocessing models to the measured lags  (Section~\ref{models}), including testing for whether a truncated disc is necessary to explain the lags.

\begin{figure}
    \includegraphics[width=\columnwidth]{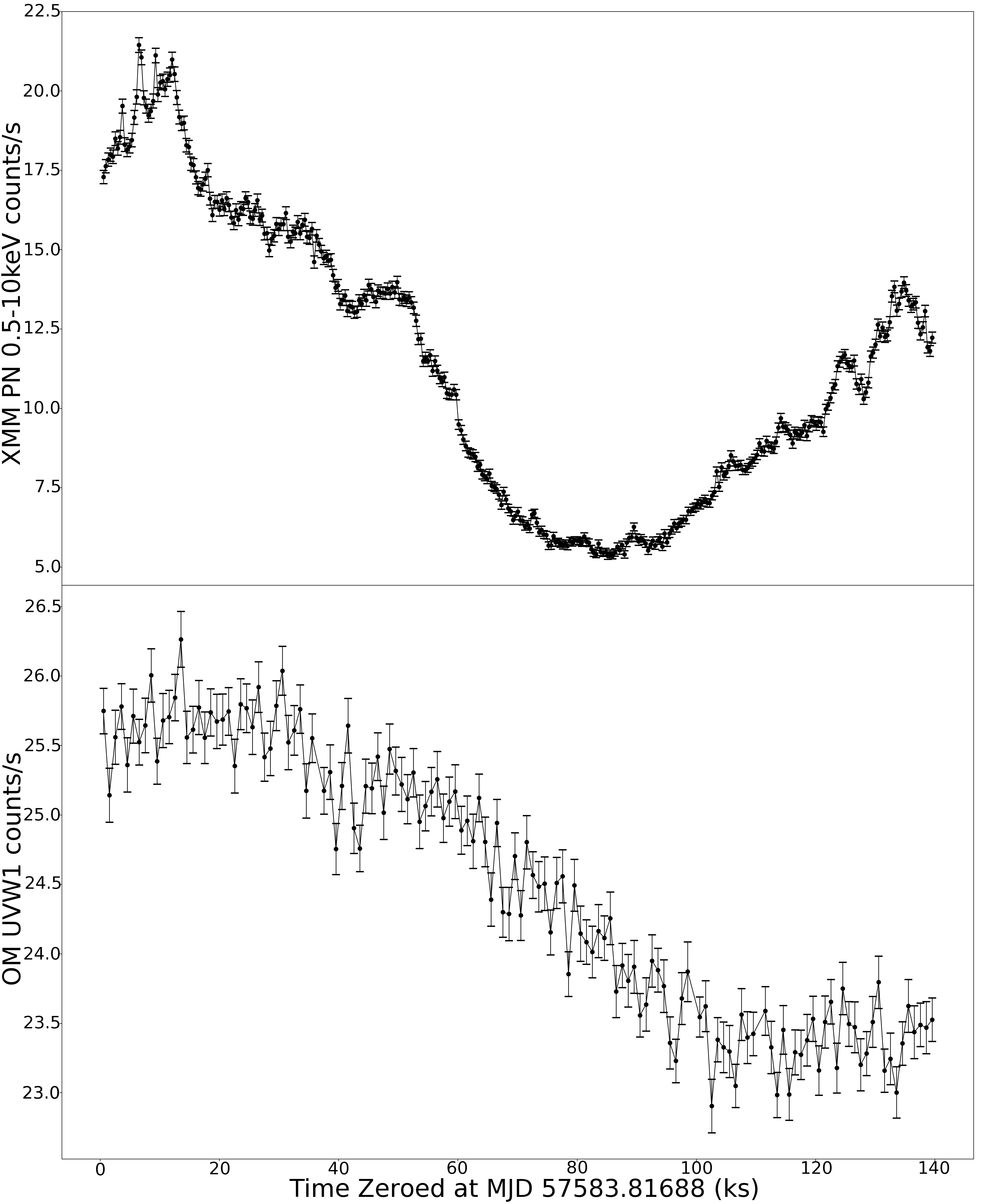}
    \caption{XMM EPIC PN 0.5-10keV X-ray light curve for NGC 4593, binned to 400s, and the XMM OM UVW1 light curve for NGC 4593, binned to 1000s.}
    \label{fig:XMMLC}
\end{figure}

\begin{figure}
    \includegraphics[width=\columnwidth]{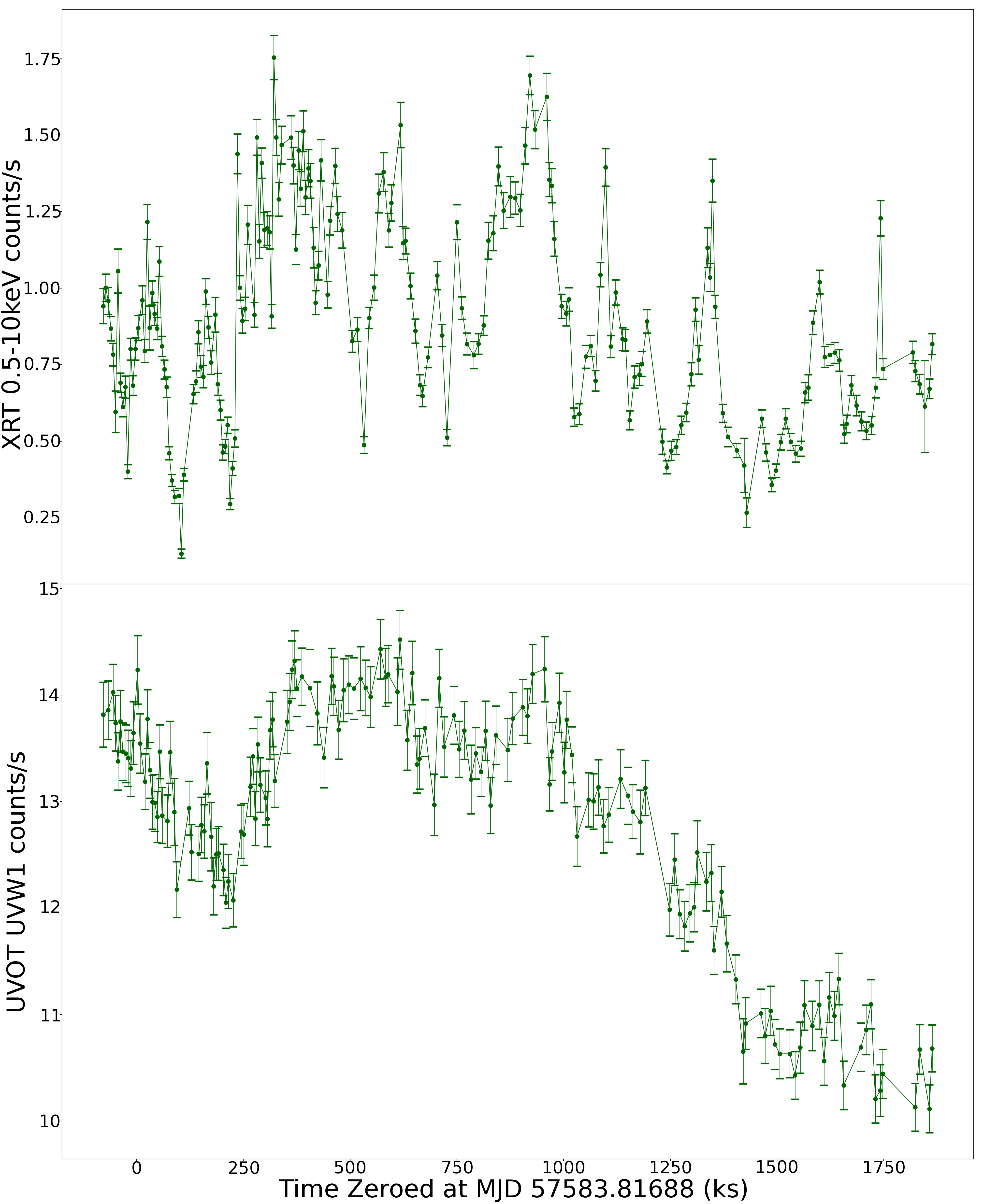}
    \caption{\swift XRT 0.5-10keV X-ray and UVOT UVW1 light curves for NGC 4593.}
    \label{fig:Swift_LCs}
\end{figure}

\begin{figure}
    \includegraphics[width=\columnwidth]{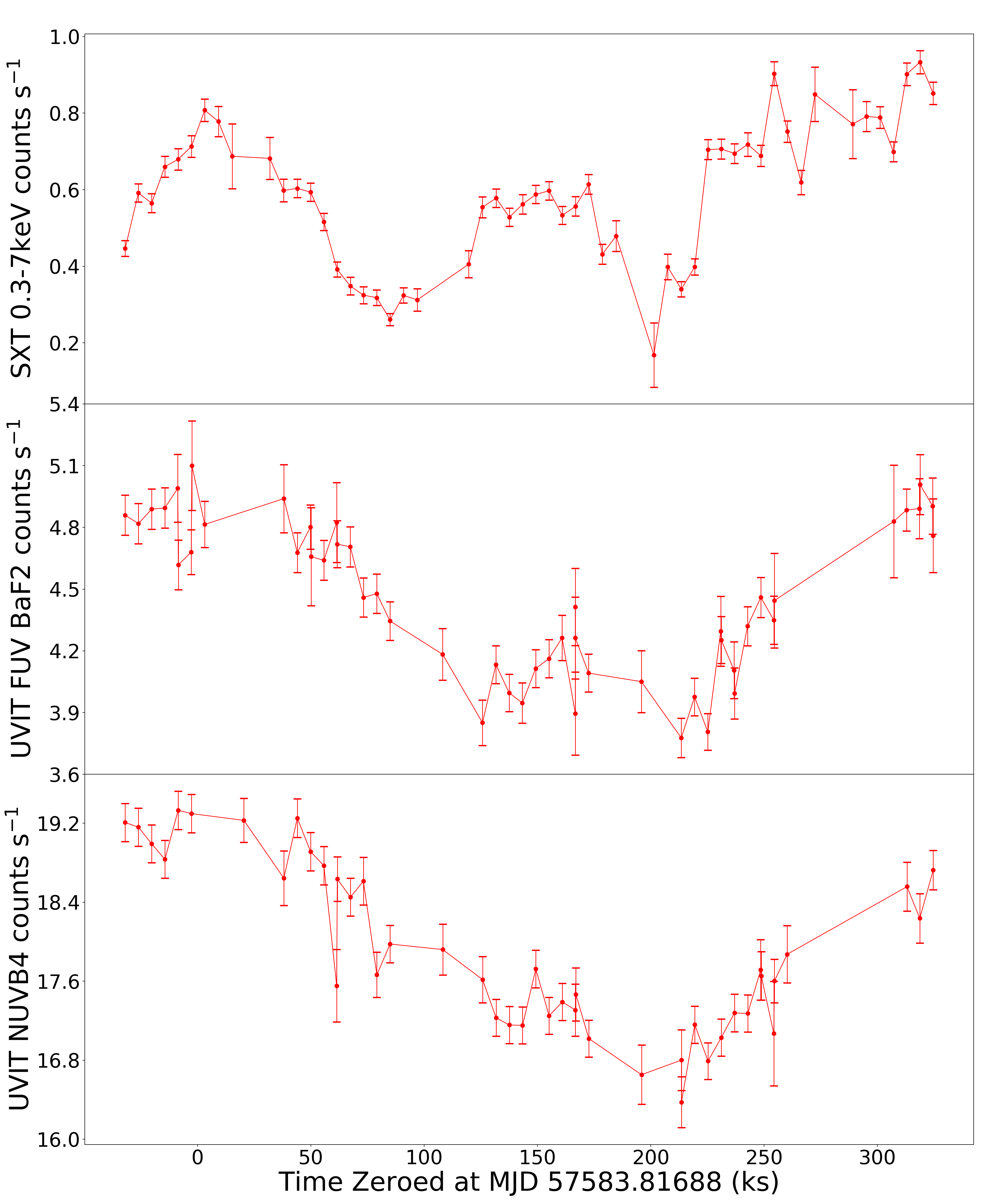}
    \caption{\astrosat SXT 0.3-7keV X-ray, UVIT FUV BaF2, and UVIT NUVB4 light curves for NGC 4593.}
    \label{fig:AstroSat_LCs}
\end{figure}

\section{Observations} \label{obs}

\subsection{\xmm Observations}
NGC~4593 was observed continuously by \xmm for one orbit of 140ks duration. The EPIC PN X-ray camera observed in small window mode and the Optical Monitor (OM) observed in Fast Mode with the UVW1 ($\lambda_{eff}=$ 2910\AA) filter. The advantage of Fast Mode is that readout is continuous with a 500ms integration time which can be binned up as required.

In Imaging Mode the shortest integration time is 800s with 300s gap for readout. A potential disadvantage of Fast Mode is that the target must be fully located within the 10$\times$10 arcsec detector window. However images of the window can be made with longer integrations and, in our case, the target remained fully located within the window. \par 

\xmm data reduction is done using the ESA Science Analysis System (SAS). From the Observation Data Format (ODF) files, SAS can generate photon event files from which we obtain light curves for the PN and OM. Initially we chose 100s bins, which we later increased. \par

In addition we ran the light curves through a filter to remove outliers more than 3$\sigma$ from nearby points, particularly a large flare in the OM that was not replicated in the X-rays, and a large flare in the X-rays that was not filtered out by the SAS pipeline's filtering process. \par

The 0.5-10 keV X-ray and UVW1 light curves, binned to 400s and 1ks respectively, are shown in Fig.~\ref{fig:XMMLC}. By eye it can be seen that both light curves have a similar shape but the variations are not simultaneous. For all light curves used in this paper we take the start of the \xmm light curves as the time zero point.

\subsection{\swift Observations}

We use the \textit{Swift} data that has already been published by \cite{mchardy_et_al_2018}, to whom we refer for observational details. The timing information which is relevant to the present paper is that NGC~4593 was observed for 1~ks nearly every orbit (96 min) from 13 to 18 July 2016 (6.4 days) and every 3 orbits for another 16.2 days. These observations are made using the X-ray Telescope (XRT) in the 0.5-10 keV band and the UV and Optical Telescope (UVOT) in 3 UV (\textit{UVW2}, \textit{UVM2}, and \textit{UVW1}) and 3 optical (\textit{U}, \textit{B}, and \textit{V}) filters. The effective wavelengths for the 6 UVOT bands in the same order as above are 1928\AA, 2246\AA, 2600\AA, 3465\AA, 4392\AA, and 5468\AA. For comparison with the \xmm OM data the most relevant UVOT filter here is \textit{UVW1}. \par

The \swift X-ray and \textit{UVW1} light curves are shown in Fig.~\ref{fig:Swift_LCs}.

\subsection{\astrosat Observations}

\astrosat observed NGC~4593 from 14 to 18 July 2016 (4.13 days). Observations were made using the Soft X-ray Telescope  \citep[SXT][]{singh_2016, singh_2017} in the 0.3-7 keV X-rays and the two cameras of the Ultra-Violet Imaging Telescope (UVIT; \cite{tandon_2017}, \cite{tandon_2020}) in the near and far UV. The filters used were the FUV BaF2 filter ($\lambda_{eff} = $1541\AA) and the NUV B4 filter ($\lambda_{eff} = $2632\AA). Note that though SXT's designed energy range is 0.3-8keV, an instrumental background line at $\sim$7.3keV reduces the useful bandwidth. \par

\astrosat has a similar low-earth orbit to \swift. Although it observed NGC~4593 continuously, earth occultation reduced the on-source time to $\sim1000-1500$s per orbit, leaving gaps of $\sim4000$s. Here we binned each observing section into one data point giving a light curve with accurately timed data points and
$\sim5600$s sampling. \par

This \astrosat data will also be shown by Kumari et al. (2022, in prep). The light curves for the binned X-ray, FUV BaF2 and NUV B4 are shown in Fig.~\ref{fig:AstroSat_LCs}.

\section{Lag Determination} \label{method}

We determined lags using the Flux Randomisation (FR) Cross-Correlation Function (CCF) method of \cite{peterson_1998}. We did not add Random Subset Selection (RSS) as it does not use the full datasets and hence artificially increases the uncertainties. However the central lag measurements do not change if RSS is added.

Another common method of lag determination is the Damped Random Walk modelling method of JAVELIN \citep{zu_2013_is}. However with only one large variation visible in the \xmm observations, the light curves are too short for JAVELIN to produce a sensible result. Thus although JAVELIN produces more precise lags, here we use only FR CCF method.

\section{Results and Analysis} \label{Analysis}

\subsection{\xmm Lag Results}

\begin{figure}
    \includegraphics[width=\columnwidth]{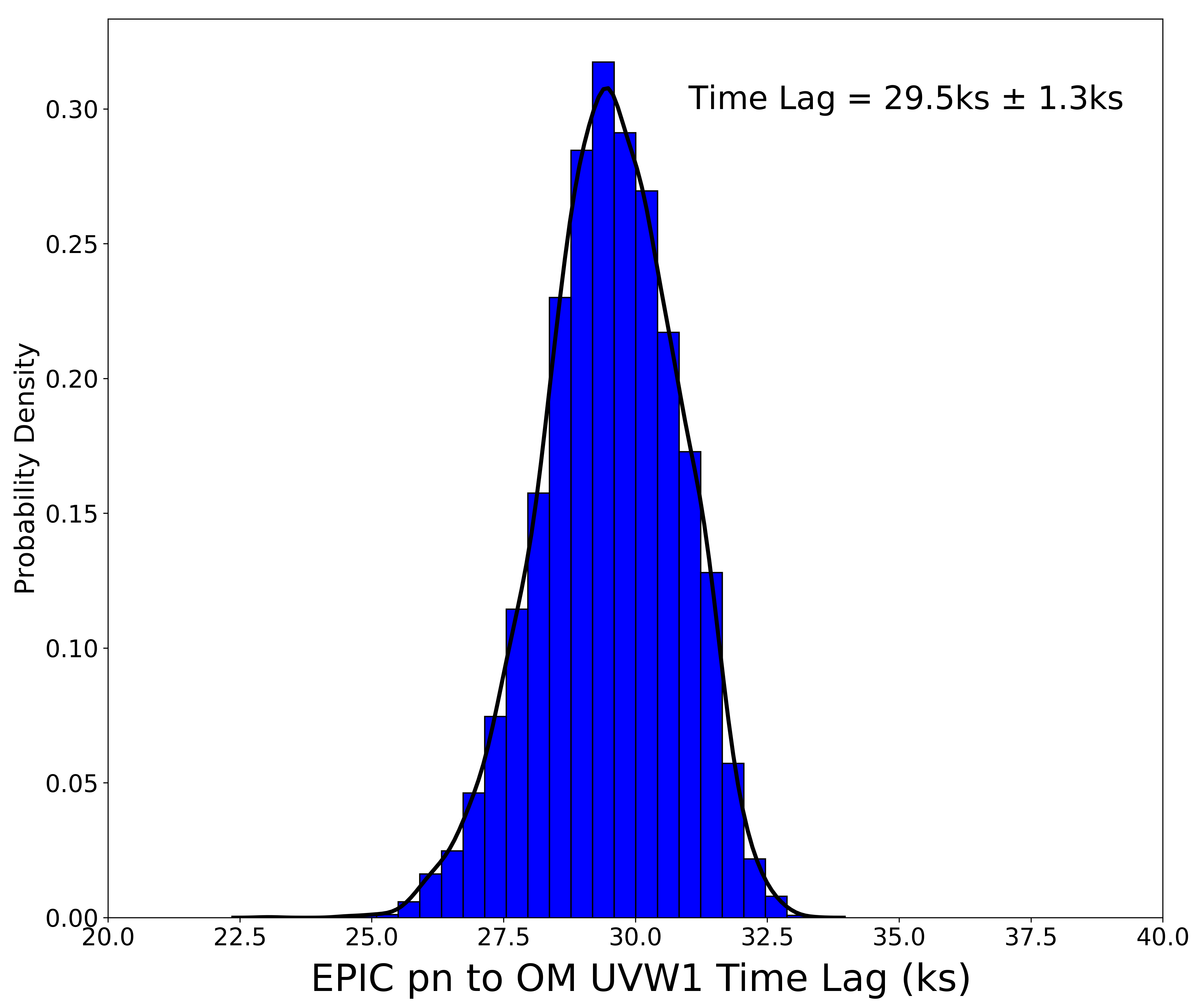}
    \caption{XMM EPIC pn to OM UVW1 Lag Distribution determined from our FR CCF trials, with a centroid lag of $29.5 \pm 1.3$ ks.}
    \label{fig:CCF}
\end{figure}

The FR CCF lag distribution between the PN X-ray and OM UVW1 lightcurves is shown in Fig.~\ref{fig:CCF}. This clean distribution has no secondary peaks and gives a centroid lag value of 29.5 $\pm$ 1.3 ks.

\begin{figure}
    \includegraphics[width=\columnwidth]{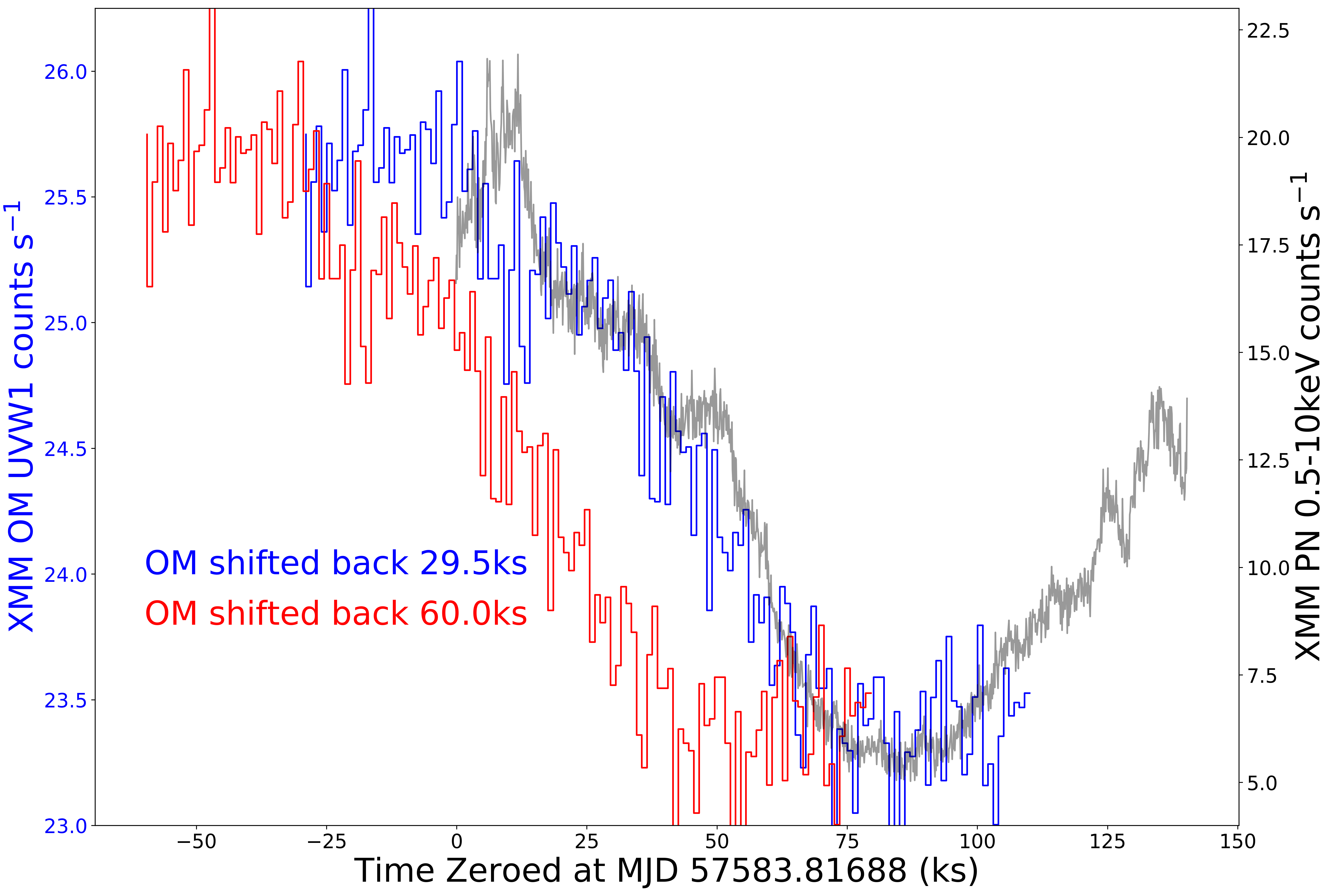}
    \caption{\xmm X-ray PN 0.5-10keV light curve with OM UVW1 light curve shifted by the \xmm (blue) and \swift (red) determined lag values.}
    \label{fig:Shifted}
\end{figure}

As reassurance that this lag is correct and not a result of a strange artefact in one or both of the light curves, in Fig.~\ref{fig:Shifted} we plot the X-ray light curve (in red) with the UVW1 light curve (in blue), scaled to the same amplitude of variability and shifted back by 29.5~ks, superposed. We can see visually that there is good agreement between the two light curves.

\subsection{\swift Lag Results}

\subsubsection{Reanalysis of CCF lags}

Comparing the above newly determined lag to that found in \cite{mchardy_et_al_2018}, we see that there is a large disparity. The value of $\sim$30ks determined from the \xmm data is less than half than that determined from the \swift data ($\sim$60ks). \par

The \xmm\, X-ray data has a much higher S/N than that from \swift, with a mean count rate $>10\times$ higher. Although \swift observed over a longer period (see Sec.~\ref{obs}), the observations, unlike with \xmm, were not continuous. Thus the \xmm\, provides 
much better temporal sampling of short-timescale variability than \swift. However it is hard to see why the improved sampling should have quite such a large effect so we reexamined the \swift data to see if we could discover the answer. \par

Upon performing the same FR CCF lag determination for the \swift XRT and UVOT UVW1 we discovered a similar lag to the one in \cite{mchardy_et_al_2018}, i.e. $\sim$60 ks. If we also shift the OM data back by this amount (Fig.~\ref{fig:Shifted}) we can see that this lag does not fit our \xmm\, data at all.

\begin{figure}
    \includegraphics[width=\columnwidth]{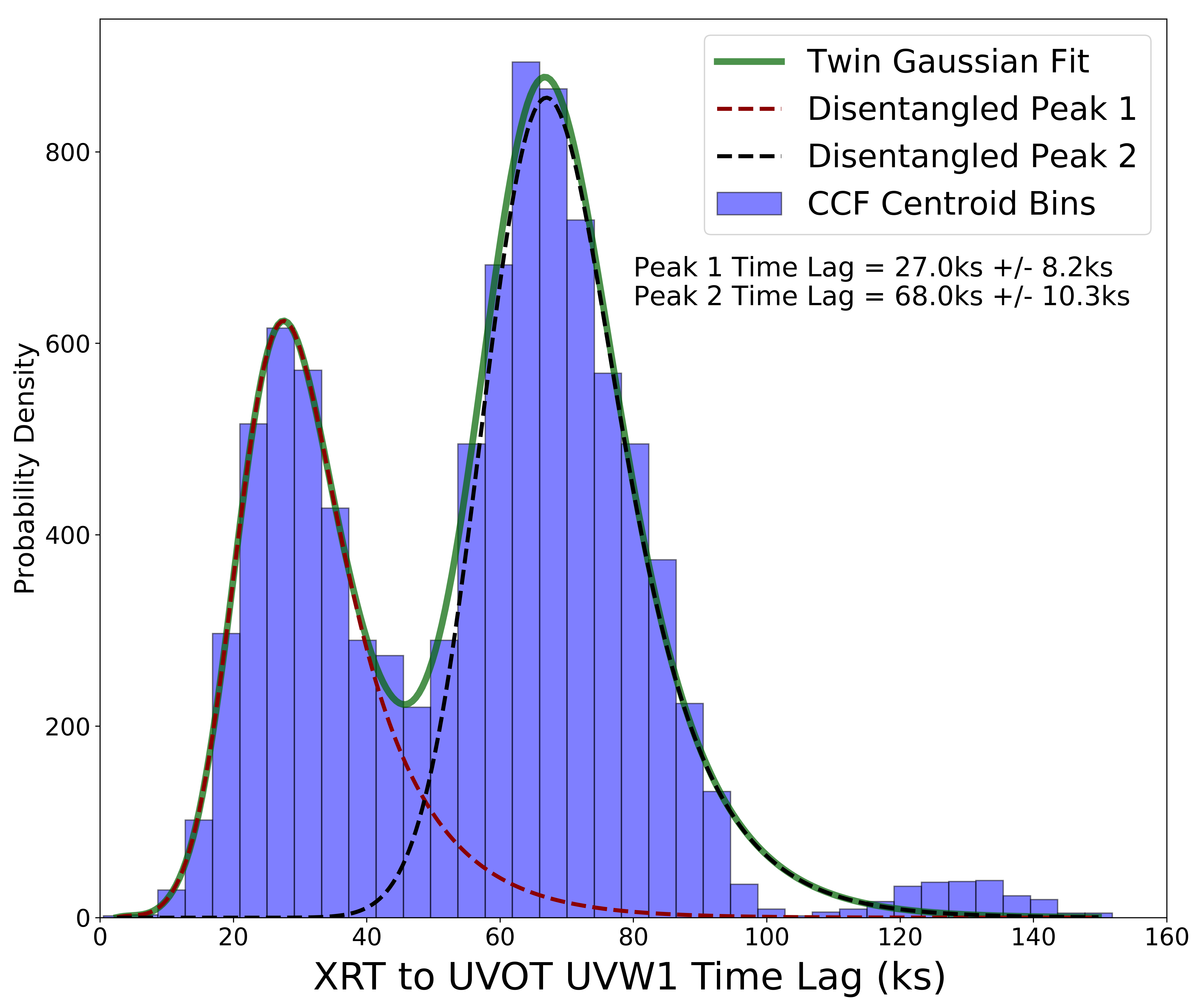}
    \caption{\swift XRT 0.5-10keV to UVOT UVW1 raw lightcurve Cross Correlation Lag Distribution with disentangled peaks.}
    \label{fig:SwiftFullLC_lag}
\end{figure}

However upon examining the Swift lag distribution, it is clear that it's bimodal, showing one peak around 30 ks and another around 70 ks (Fig.~\ref{fig:SwiftFullLC_lag}). By fitting a convolution of two asymmetric, decaying gaussian functions to this bimodal distribution using scipy's \texttt{curve\_fit}, we can then disentangle them to retrieve the lags for each peak.

The lags we retrieve from these disentangled distributions are 27.0 $\pm$ 8.2 ks and 68.0 $\pm$ 10.3 ks.

The mean level of X-ray light curve does not change noticeably over the period of the observations but the mean UV level decreases. To test the idea that there are actually two lags within the \swift data, we produce a smoothed UV light curve which we then subtract from the original light curve to produce a detrended light curve. We can then measure lags using both the smoothed and detrended UV light curves. \par

\subsubsection{Detrending with LOWESS}

To do the smoothing we use a method called Locally Weighted Scatterplot Smoothing (LOWESS), which is a non-parametric regression fitting technique and which we used successfully previously \citep{pahari_2020}.  We use the Python module \texttt{statsmodels lowess} function. The main input of this function, other than the data itself, is the fraction of the light curve the function will smooth along. Using a very small smoothing fraction will, after detrending, leave only very short-timescale variations. Thus this fraction should be significantly larger than any timescales of interest. 

Our total light curve length is $\sim$22 days and the lags seen in the original lightcurves are on the order of a day so we investigated the effect of various smoothing timescales longer than a few days. 

\begin{figure}
    \includegraphics[width=\columnwidth]{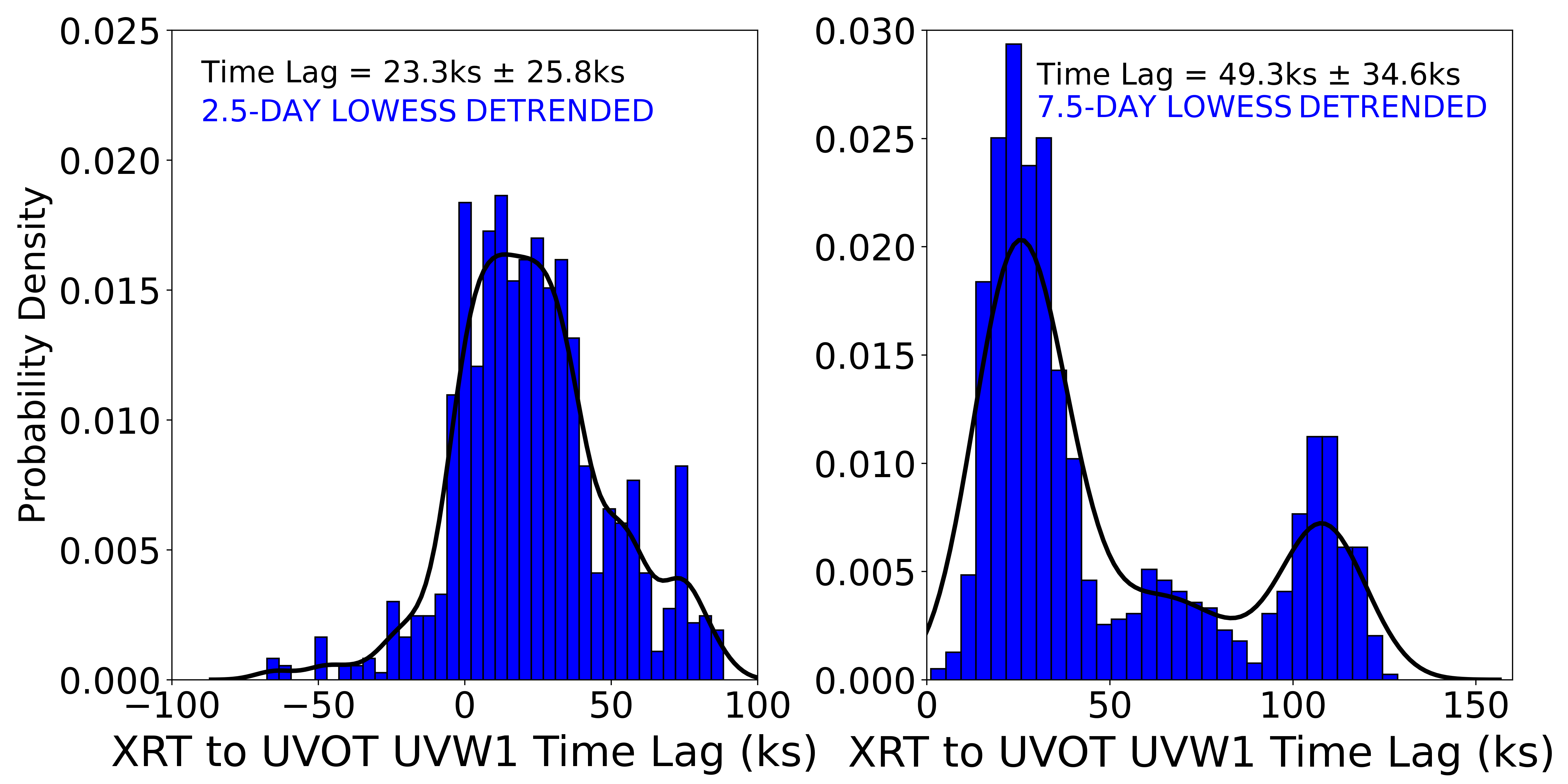}
    \caption{Cross Correlation Lag distributions from \swift un-detrended XRT 0.5-10keV to UVOT UVW1 detrended with 2.5-day (left) and 7.5-day (right) Lowess smoothing timescales.}
    \label{fig:SwiftCompSubLCs_lag}
\end{figure}

\begin{figure}
    \includegraphics[width=\columnwidth]{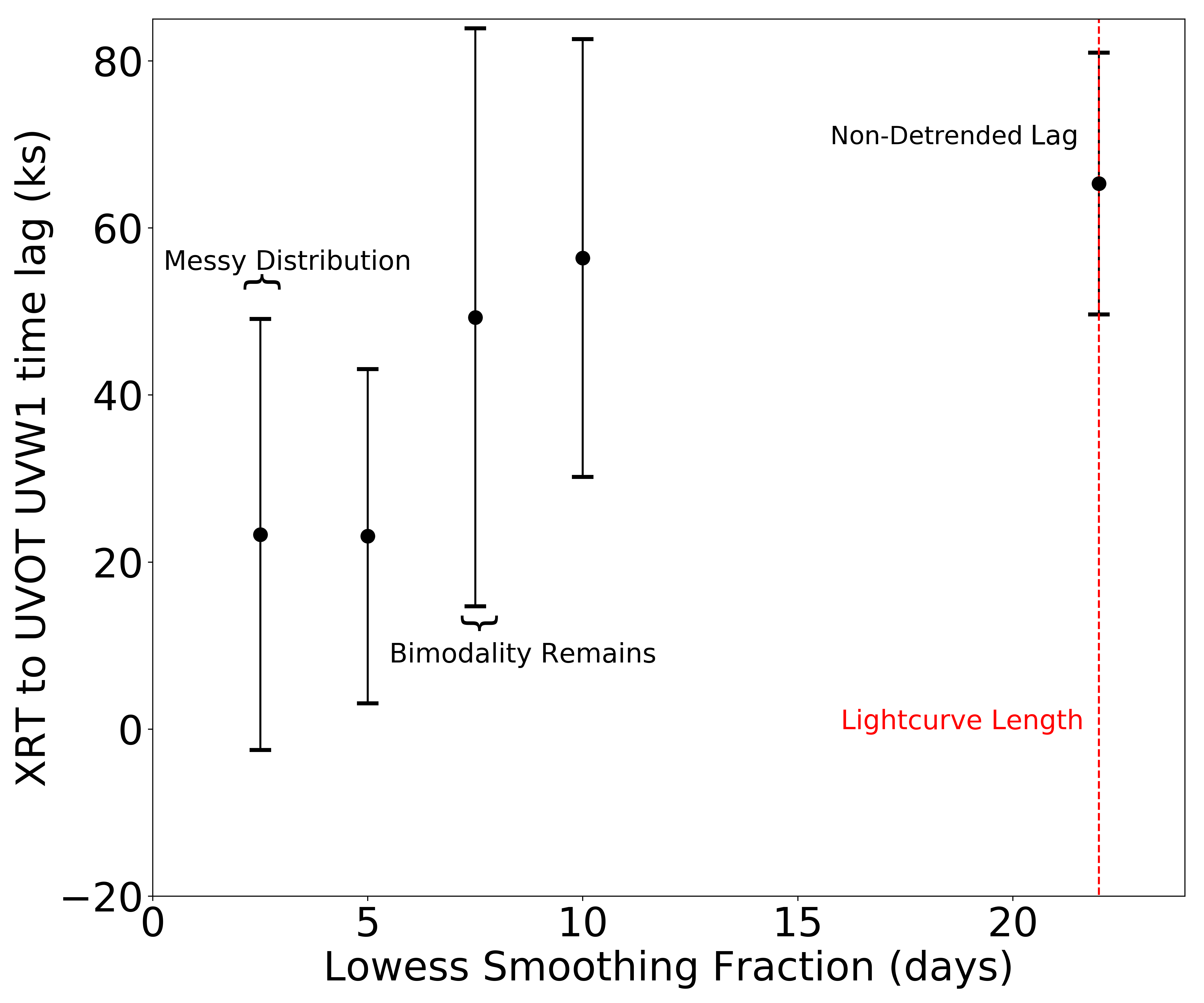}
    \caption{Lag Determinations vs Lowess Smoothing Fraction for the \swift UVOT lightcurves.}
    \label{fig:LagvsBoxcar}
\end{figure}

In the left and right panels of Fig~\ref{fig:SwiftCompSubLCs_lag} we show the lag distributions between the un-detrended Swift X-ray lightcurve and the detrended UVOT lightcurve after detrending with 2.5d and 7.5d timescales respectively.  The 7.5d timescale still leaves a large residue of the longer peak. The 2.5d detrending timescale shows a messy distribution, as the detrending is beginning to remove short-term lag features and therefore reducing the confidence of the lag determination, as we are approaching the timescale of the lags themselves. The resulting lag measurements for different detrending timescales are plotted in Fig.~\ref{fig:LagvsBoxcar}. \par

\begin{figure}
    \includegraphics[width=\columnwidth]{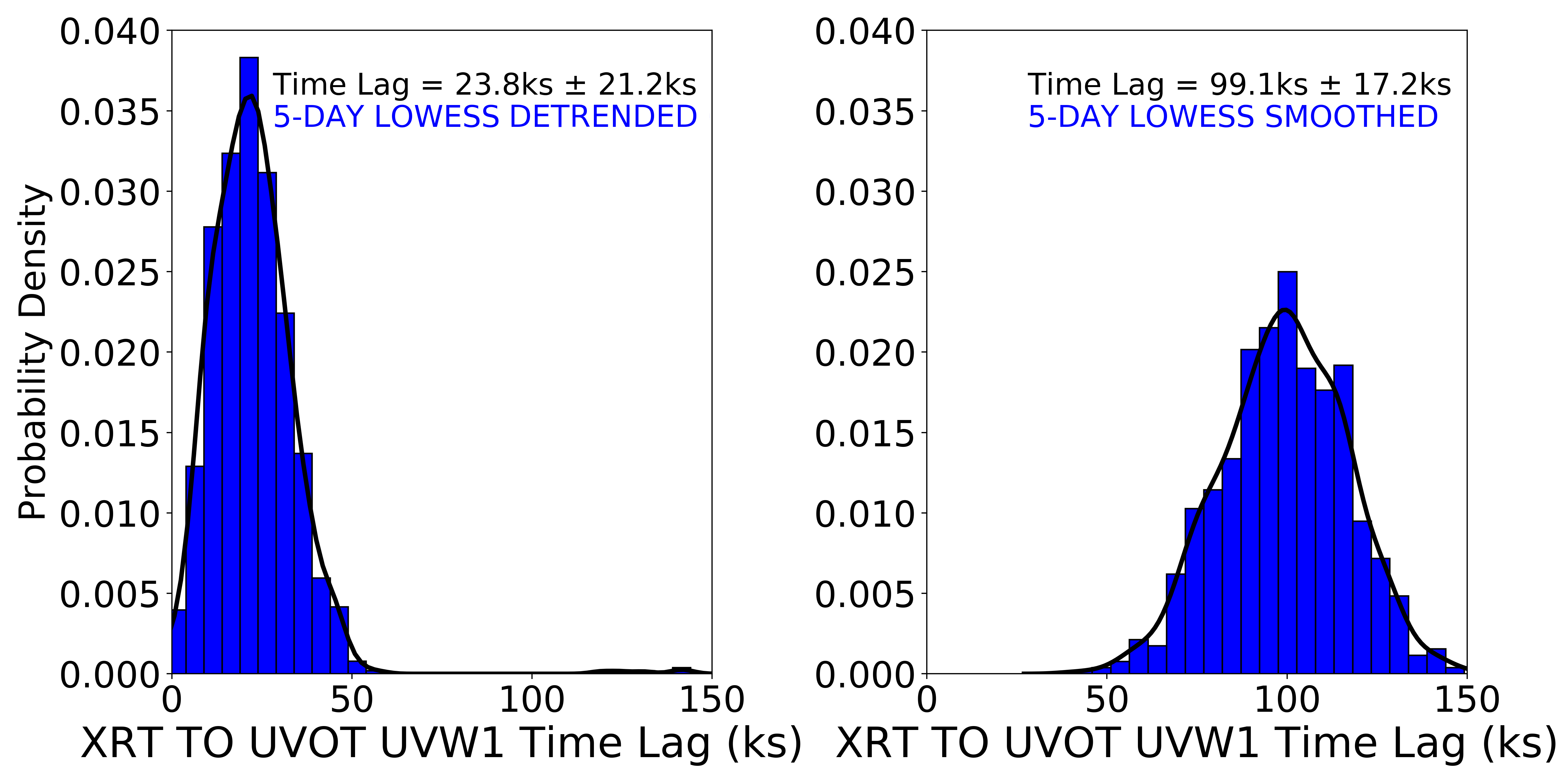}
    \caption{Cross Correlation Lag Distributions from \swift XRT 0.5-10keV to UVOT UVW1 where both lightcurves have been 5-day Lowess Detrended (left) and Smoothed (right), respectively.}
    \label{fig:Swift5dayLC_lags}
\end{figure}

The trend in the UV is larger than in the X-rays and detrending also in the X-rays has usually had little extra effect on the lags. However to check whether detrending the X-rays as well as the UVOT has any effect on the lags here we detrend both bands with a 5d smoothing timescale. The resultant lag distribution is shown in Fig.~\ref{fig:Swift5dayLC_lags}, left panel, giving a lag of 23.8 $\pm$ 21.2ks, which is almost identical to that where only the UVOT was detrended.

\begin{figure}
    \includegraphics[width=\columnwidth]{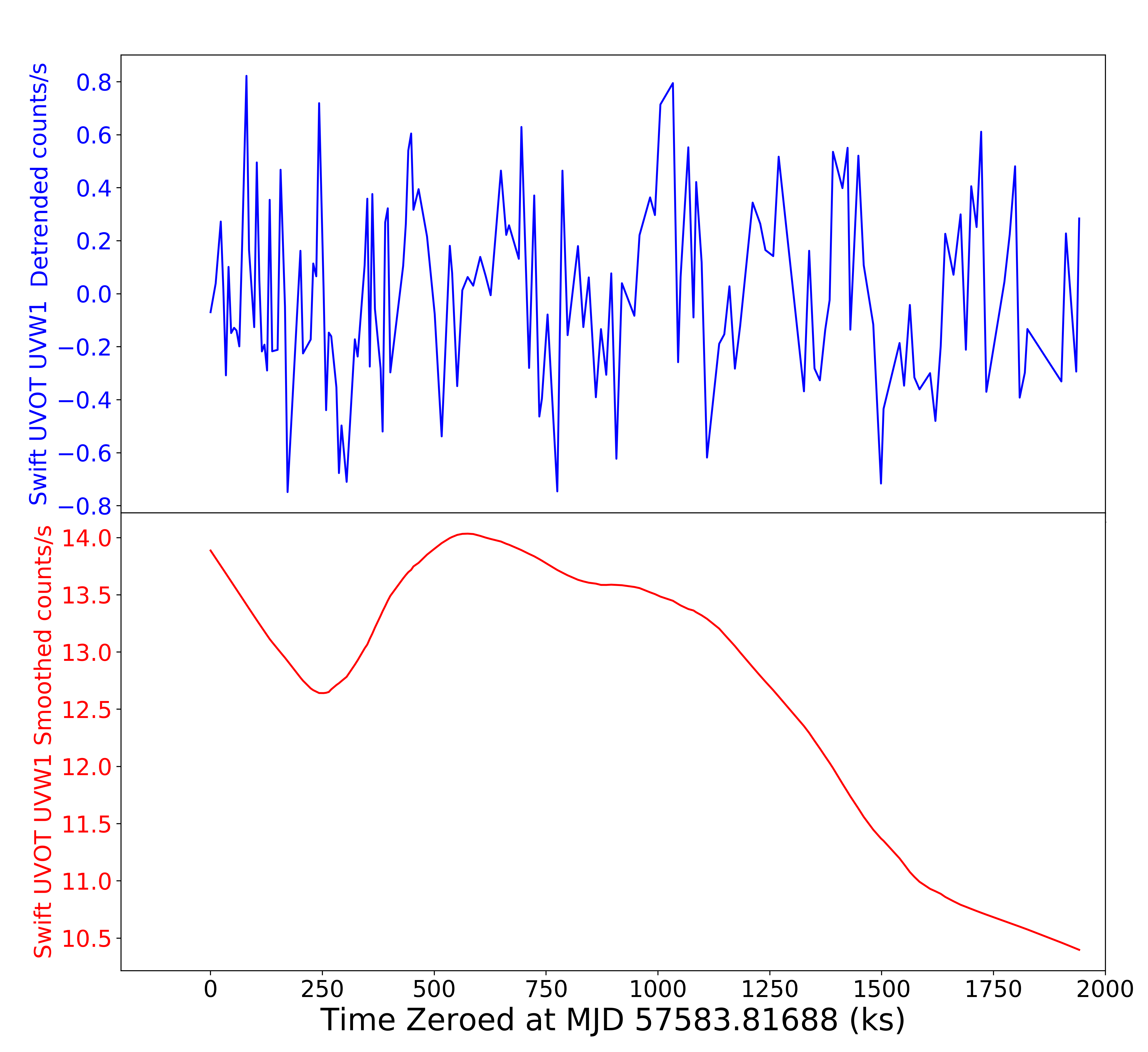}
    \caption{\swift UVOT UVW1 5-day Lowess Detrended and Smoothed Light Curves.}
    \label{fig:SwiftLowessLCs}
\end{figure}

To determine whether the longer (68ks) lag found in the raw, un-detrended UVOT and X-ray lightcurves (Fig.~\ref{fig:SwiftFullLC_lag}) might result from the slower variations which we have subtracted from the detrended lightcurves, we performed a lag measurement using the smoothed X-ray and smoothed UVW1 light curves. The detrended and smoothed UVOT lightcurves are show in Fig.~\ref{fig:SwiftLowessLCs} and the lag distribution is shown in Fig.~\ref{fig:Swift5dayLC_lags}, right panel. The resultant lag is 99.1 $\pm$ 17.2ks, but with no sign of a peak in the 20-30 ks range. This lag probably represents the longer lag seen in the full light curves. \par

\subsubsection{Testing alternate explanations}

To check whether the different lags were the result of energy dependent lag components, we carried out the above detrending analysis separately on the soft (0.5-2keV) and hard (2-10keV) \swift light curves but found the same result.

\begin{figure}
    \includegraphics[width=\columnwidth]{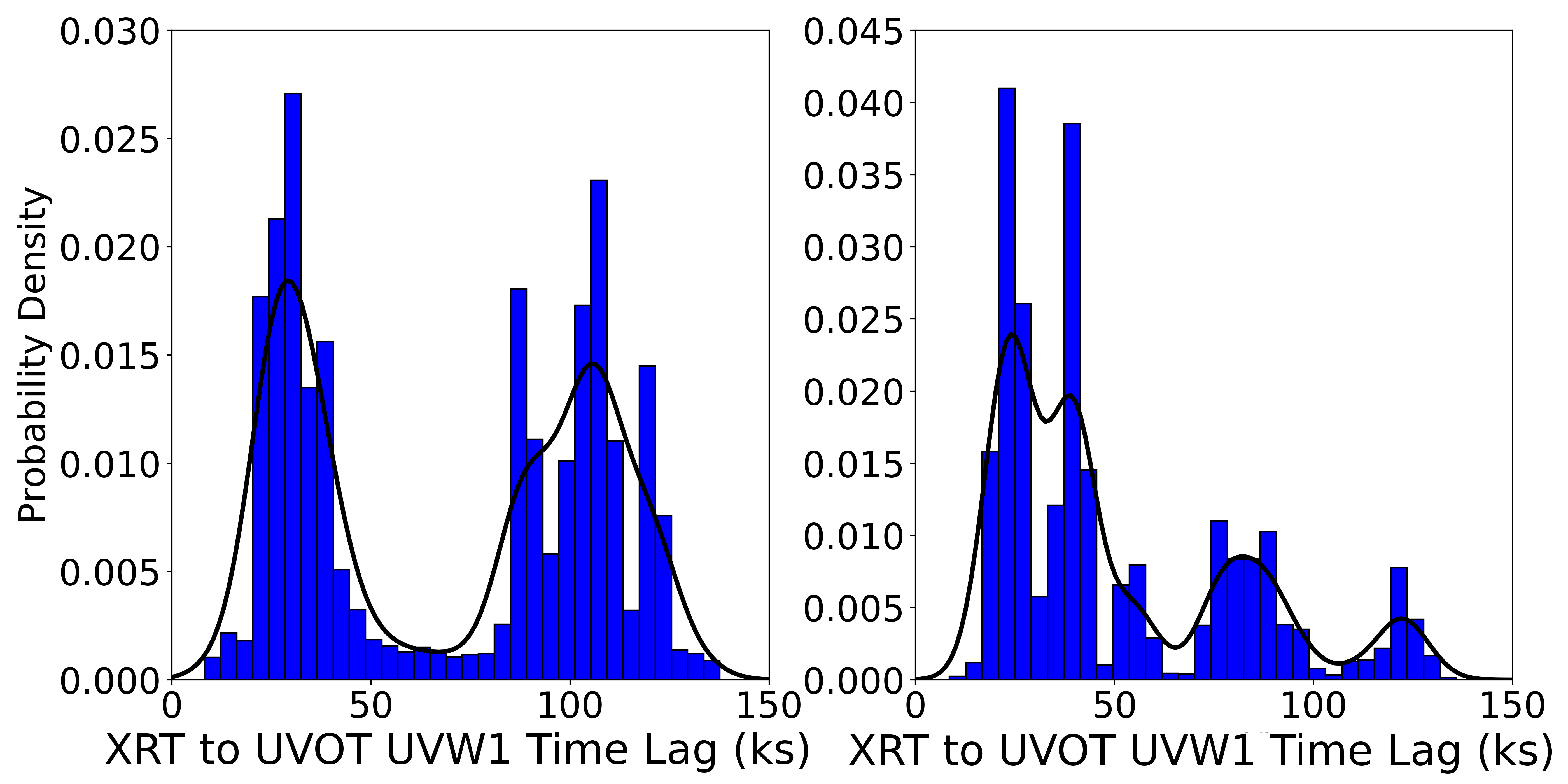}
    \caption{\swift XRT 0.5-10keV to UVOT UVW1 High Intensity Half only (left) and Low Intensity Half (right) CCF Lag Distributions.}
    \label{fig:Swift_IntensityLags}
\end{figure}

In addition to check that the two peaks are not related to the sampling pattern, e.g. with the initial high sampling intensity period of 6.4 days producing a short lag and the remaining 16.2d of lower intensity sampling producing a longer lag, we carried out a separate lag analysis for both periods (Fig.~\ref{fig:Swift_IntensityLags}). Both peaks are present in both sections although the reduction in data leads to messier distributions. This analysis shows that the \swift data does contain a short lag consistent with that seen in the XMM data. There is a small caveat that although the \swift UVOT and XMM OM are very similar instruments with similar filters, there has been some loss of UV throughput in the OM. This loss has resulted in the central wavelength of the UVW1 filter now being $\sim$2910\si{\angstrom} whereas that of \swift is $\sim$2600\si{\angstrom}. This small wavelength difference may result in a similarly small lag difference. However the uncertainty in the \swift lag is too large for us to detect that difference. \par

To check whether the bimodal lags from the raw light curves were a result of stray points resulting from minor irregularities in the data reduction process, we compared these light curves, generated with an in-house University of Southampton pipeline for \cite{mchardy_et_al_2018}, to lightcurves from the same observation period generated with the University of Leicester's online product builder \footnote{\url{https://www.swift.ac.uk/user_objects/index.php}}. The lag distributions we determined for these regenerated light curves were identical to those found with our in-house light curves for both the raw, detrended, and smoothed light curves, showing that the distributions we detect are therefore not the result of any issue in the reduction process.

\begin{figure}
    \includegraphics[width=\columnwidth]{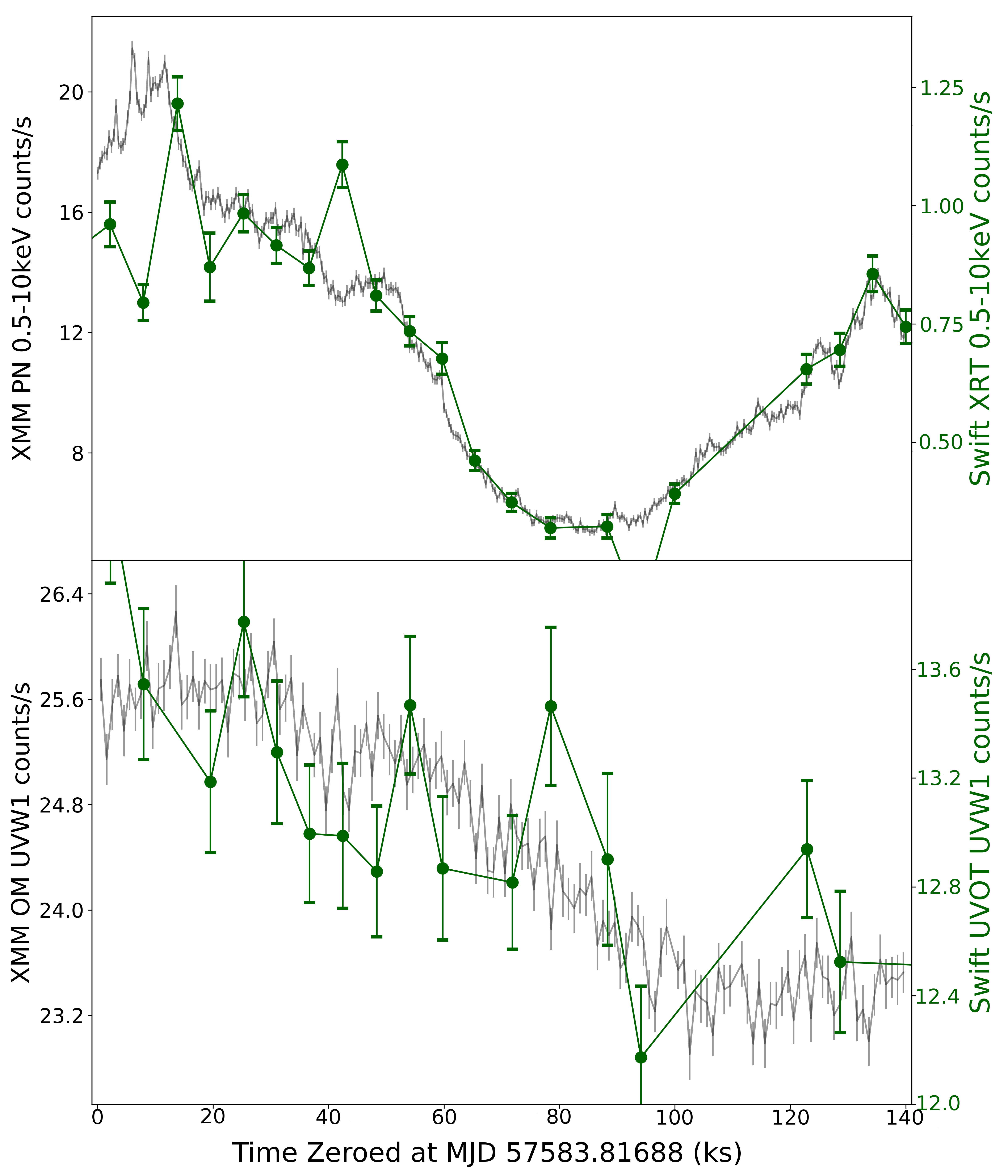}
    \caption{\swift XRT 0.5-10keV and UVOT UVW1 plotted along with the XMM EPIC pn and OM UVW1 light curves.}
    \label{fig:SwiftvsXMM}
\end{figure}

The \swift observation period overlaps entirely with that of XMM. The XMM and \swift X-ray and UV light curves are plotted together in Fig.~\ref{fig:SwiftvsXMM}, showing good agreement, particularly in the X-rays where S/N is greatest. The much better sampling of the XMM data in both bands is apparent.

\begin{figure}
    \includegraphics[width=\columnwidth]{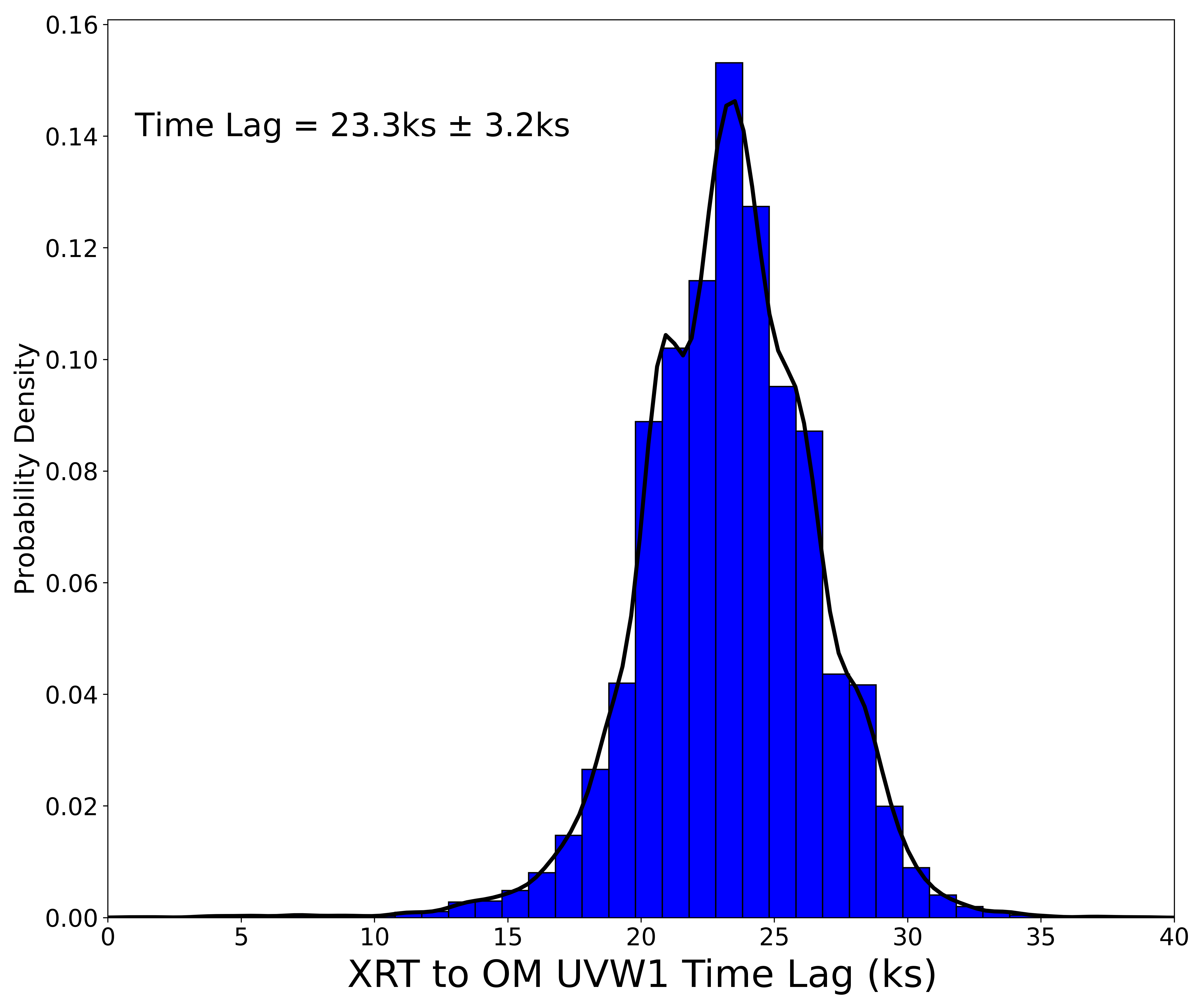}
    \caption{\swift XRT 0.5-10keV to XMM OM UVW1 Cross Correlation Time Lag.}
    \label{fig:SwifttoOM_lag}
\end{figure}

As a check, we can perform a lag analysis of the \swift XRT light curve with the XMM OM. The resulting lag distribution is shown in Fig.~\ref{fig:SwifttoOM_lag} and the measured lag is 23.3 $\pm$ 3.2 ks with no other peaks at larger lags. This lag is very similar to that provided by both the XMM data on its own and by the detrended \swift data. 
This result implies that the detrended \swift lag correctly represents the true lag that would be measured in short, well sampled, light curves. \par 

\subsection{\astrosat Lag Results}

As a further diagnostic of the short timescale lags we can use another overlapping light curve set from \astrosat.

\begin{figure}
    \includegraphics[width=\columnwidth]{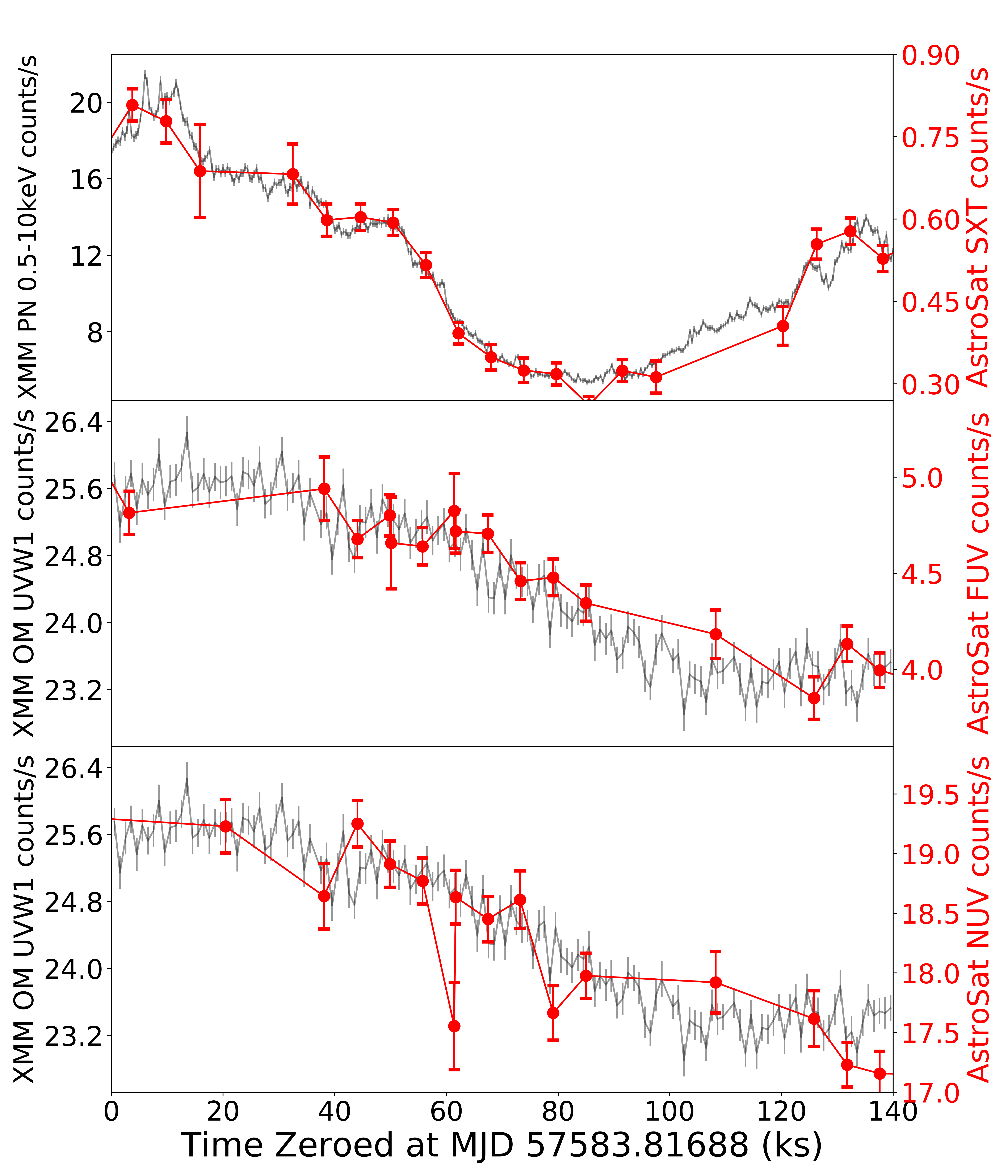}
    \caption{\astrosat SXT 0.3-7keV, FUV BaF2, and NUVB4 plotted along with the XMM EPIC pn and OM UVW1 light curves.}
    \label{fig:AstroSatvsXMM}
\end{figure}

The central wavelengths of the \astrosat NUV B4 and FUV BaF2 filters are 2632\si{\angstrom} and 1541\si{\angstrom} respectively. The central wavelength of the NUV filter is similar to that of the \swift UVOT UVW1 filter. In Fig.~\ref{fig:AstroSatvsXMM} we compare the \astrosat SXT, FUV and NUV lightcurves with the X-ray and UVW1 lightcurves from XMM. As with \swift, the \astrosat sampling does not match that of XMM but similar broad features are visible in both the X-ray and UV lightcurves.

\begin{figure}
    \includegraphics[width=\columnwidth]{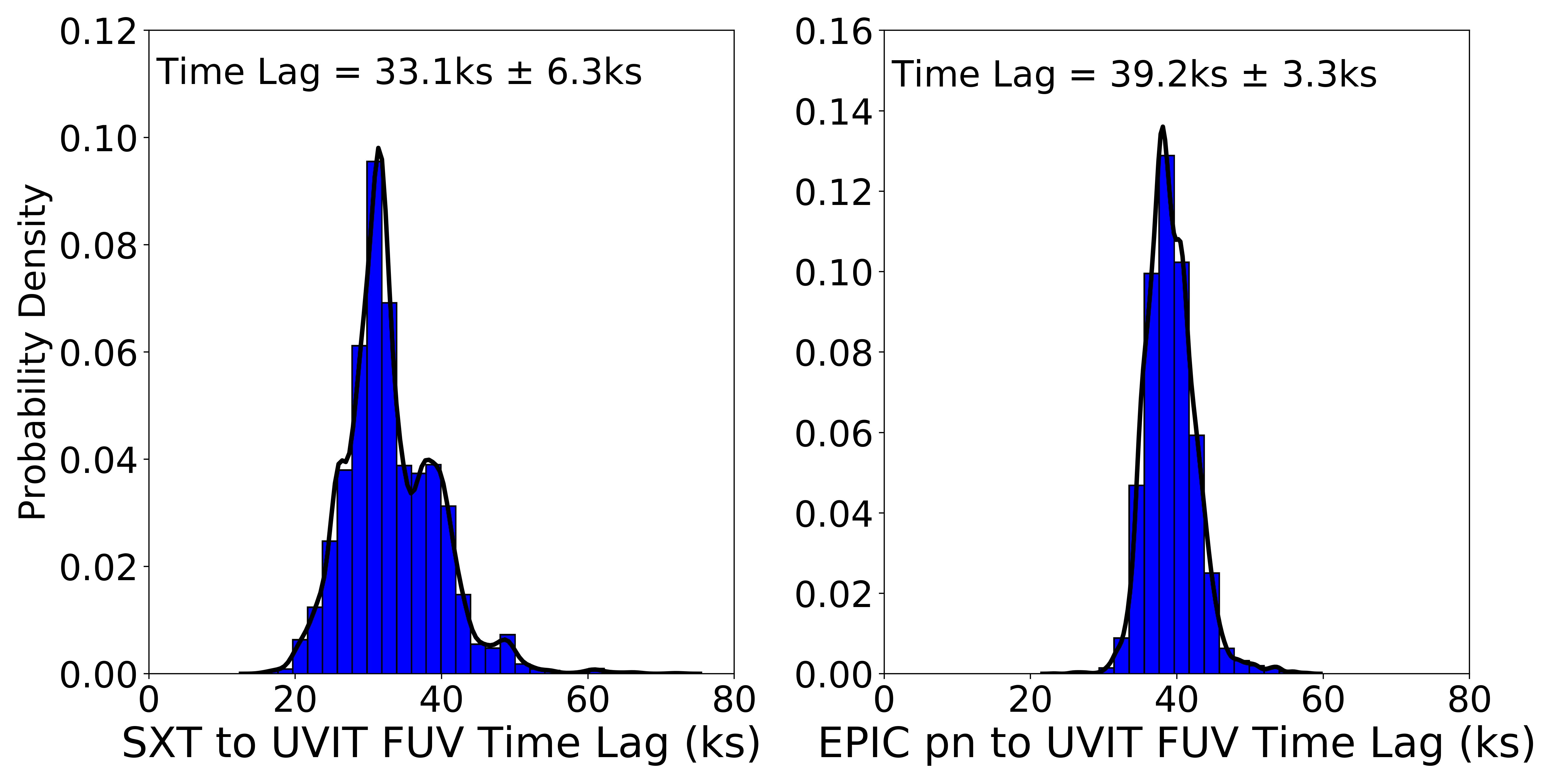}
    \caption{\astrosat SXT 0.3-7keV to UVIT FUV BaF2 (left) and XMM EPIC pn 0.5-10keV to UVIT FUV BaF2 (right) Cross Correlation Time Lags.}
    \label{fig:UnSubFUV_lags}
\end{figure}

We begin to measure the lags using the original, un-detrended data. Here in Fig.~\ref{fig:UnSubFUV_lags} the lag distribution of SXT to FUV shows a larger lag than would be expected for its short central wavelength. On the right of the same figure, we determine the FUV lag relative to the \xmm EPIC pn X-rays and similarly find a larger than expected main lag. \par

\begin{figure}
    \includegraphics[width=\columnwidth]{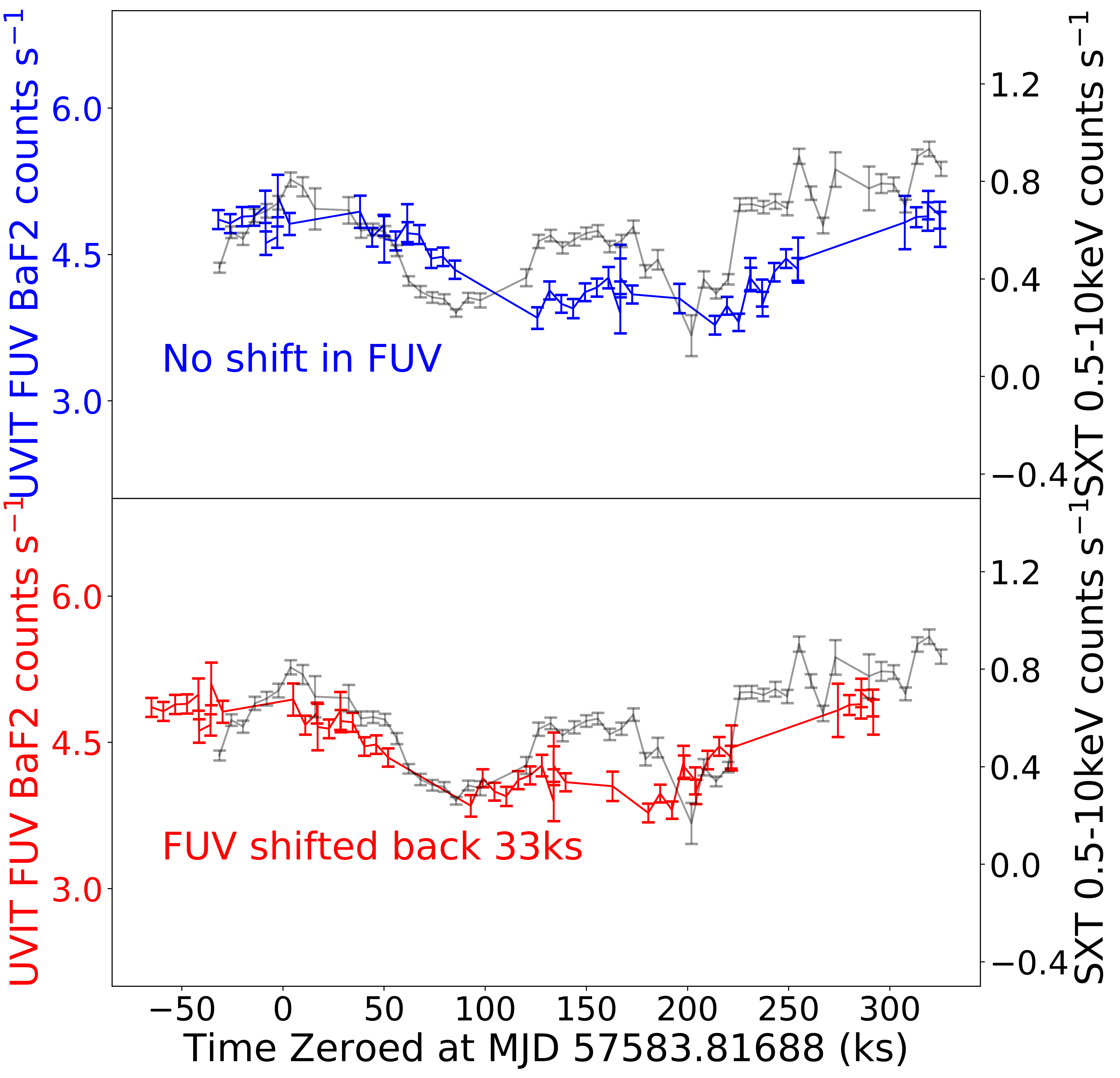}
    \caption{\astrosat SXT 0.3-7keV plotted with UVIT FUV BaF2 (top) and SXT 0.3-7keV plotted with UVIT FUV BaF2 shifted by -33ks (bottom) light curves.}
    \label{fig:FUV_Comp_Lags}
\end{figure}

If we look at the FUV data points overlaid onto the SXT data, we can see from the top panel of Fig.~\ref{fig:FUV_Comp_Lags} that the FUV visually appears to be compatible with a small lag. Additionally it can be seen from the bottom panel that applying the 33ks lag detected from the raw lightcurves (Fig.~\ref{fig:UnSubFUV_lags}, left panel) to the FUV light curve is larger and does not visually align the light curves. \par

Therefore with the evidence of long-timescale trends found in the \swift light curves, we can apply a LOWESS detrending to see if it alters the FUV lags. In this case, though the light curve is shorter than \swift, variability power can still 'leak' into a light curve if there is variability on timescales longer than the length of our observation (\cite{deeter_1982}), which according to our \swift observations there appears to be. In this case then we can perform a LOWESS detrending over the entire length of the light curve. Though the light curve is only $\sim$4 days long, it can be seen in Fig.~\ref{fig:LagvsBoxcar} that even an averaging window down to 2.5 days does not change the result much as it is still a timescale an order of magnitude larger than our expected lags, therefore a much larger 4 day smoothing should be sufficient for our detrending. \par

\begin{figure}
    \includegraphics[width=\columnwidth]{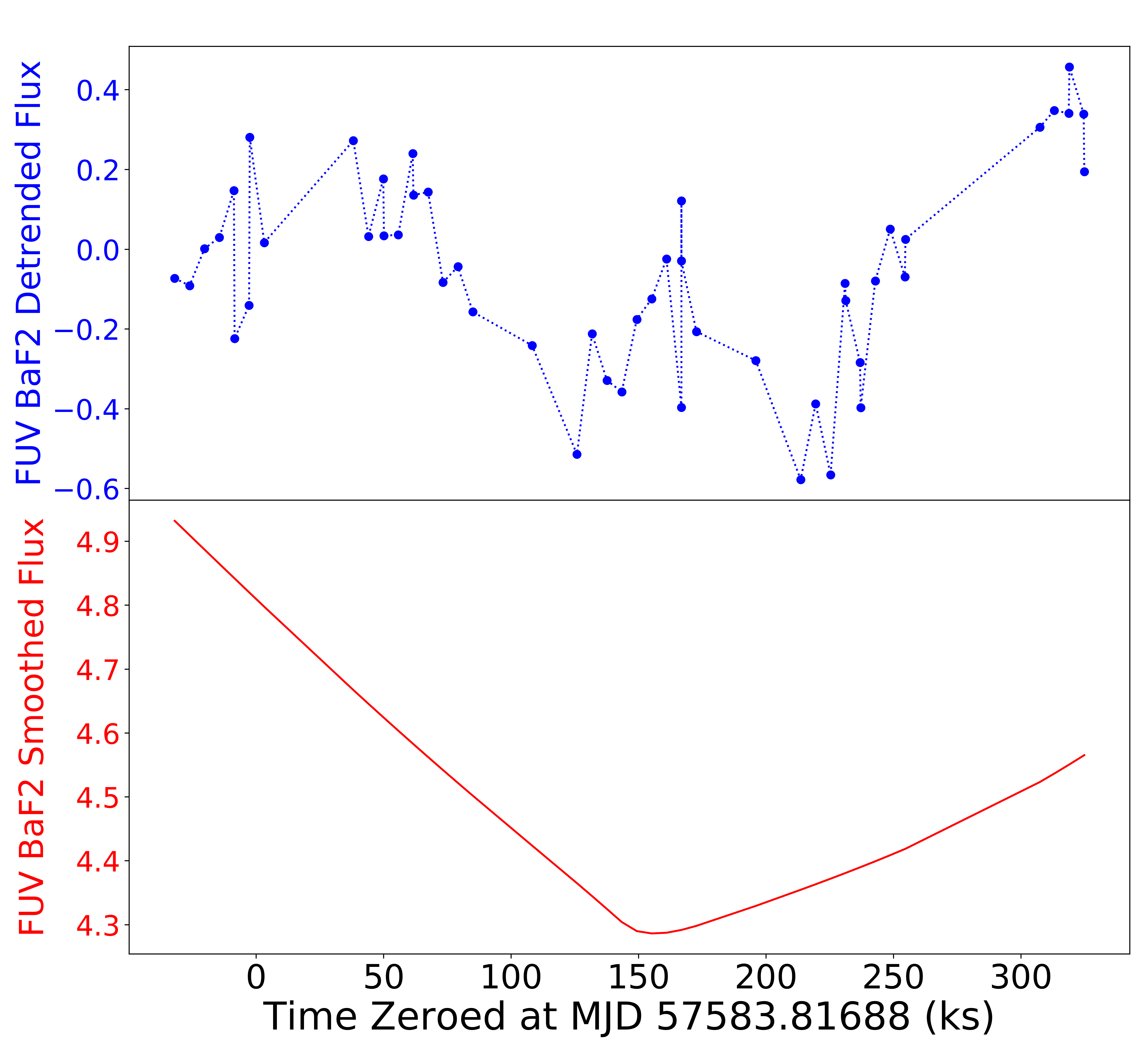}
    \caption{\astrosat UVIT FUV BaF2 4-day Lowess Detrended and Smoothed Light Curves.}
    \label{fig:FUVLowessLCs}
\end{figure}

In Fig.~\ref{fig:FUVLowessLCs} the results of the detrending can be seen. Note that this detrending detects a very similar trend to the \swift over the equivalent time period, giving more confidence that the trend being removed is physically real. \par

We note that we for the NUV B4 lag distributions, the lags we detect appear to be consistent with the rest of the lag spectrum with regards to their wavelength and Lowess detrending does not affect the detected lag. We speculate that a reason for this could be that as the NUV is a longer wavelength band, the damping effects we see at larger responding wavelengths have reduced the magnitude of the long-term contribution. This can be visually seen in the light curves, as the trend on the first 150ks of the FUV light curve results in an overall flux decrease of $\sim$18\%, whereas in the equivalent period of NUV the decrease appears to only be $\sim$8\% of the flux. Though NUV is approximately the same wavelength as \swift's UVW1, the light curve is far shorter than the \swift light curve and thus the lag determination may not be sensitive enough to detect the long-term variability's effect in such a short light curve at this waveband, which would be consistent with what we see in the even-shorter \xmm UVW1 light curves. \par

\begin{figure}
    \includegraphics[width=\columnwidth]{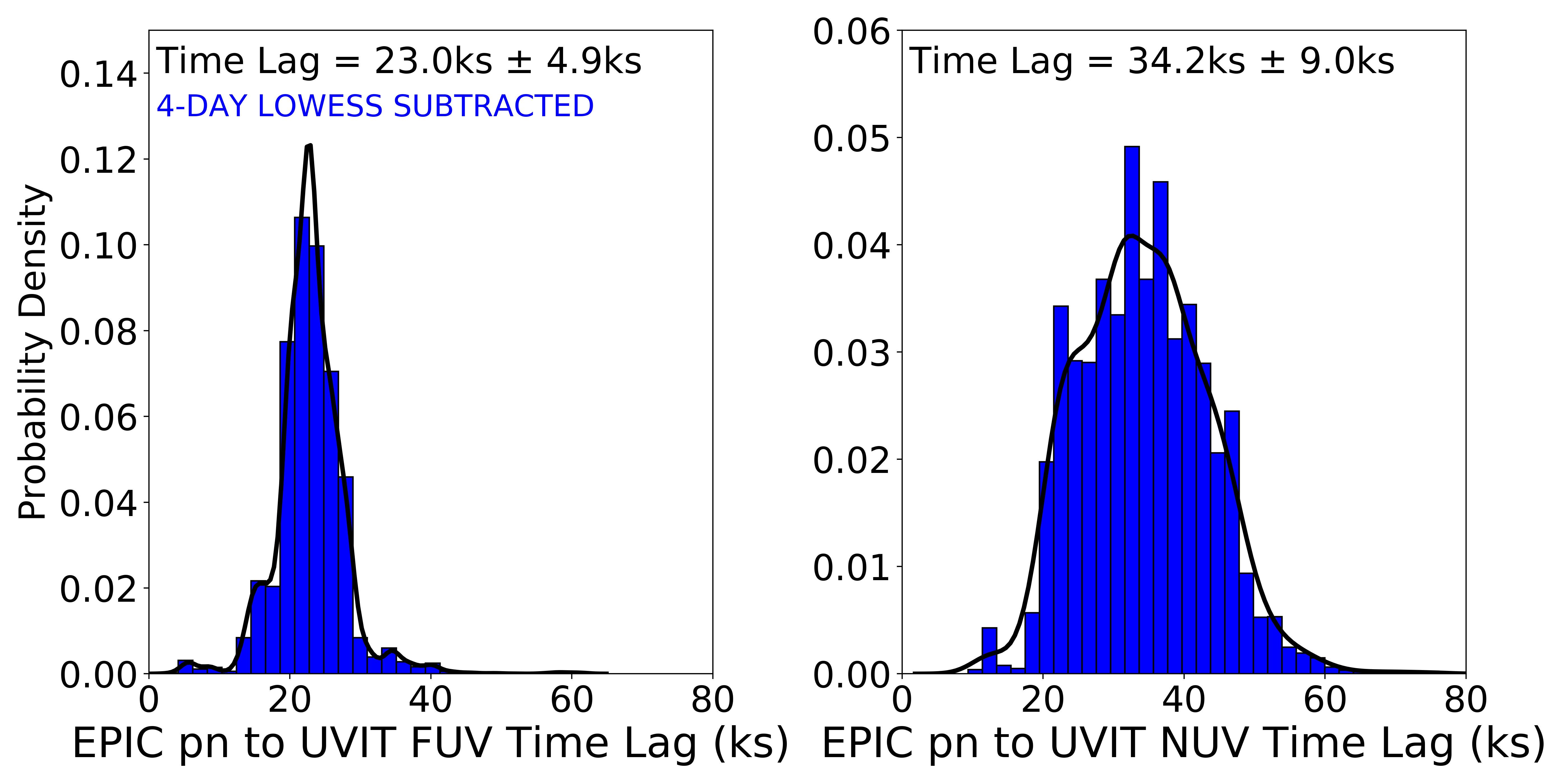}
    \caption{\astrosat SXT 0.3-7keV to Lowess-Detrended UVIT FUV BaF2 (left) and SXT 0.3-7keV to UVIT NUVB4 (right) Cross Correlation Time Lags.}
    \label{fig:SXT_lags}
\end{figure}

Therefore, the lag distributions between the \astrosat SXT 0.3-7keV X-rays \& the Lowess-Detrended FUV, and between the SXT 0.3-7keV X-rays \& the NUV are shown in Fig.~\ref{fig:SXT_lags}, finding lags of 23.0 $\pm$ 4.9 ks and 34.2 $\pm$ 9.0 ks, respectively. This FUV lag is slightly longer than might be expected for its wavelength, but as we will examine in Section \ref{models}, there may be other factors than disc reverberation in NGC 4593. It is also noted that the FUV shows physical similarities to the \swift UVW2 light curve, which also shows similar lag taken relative to UVW1's lag, so this result is consistent with the appearance of the light curve. The NUV lag is in agreement with the short timescale lags found with \swift and \xmm.

\begin{figure}
    \includegraphics[width=\columnwidth]{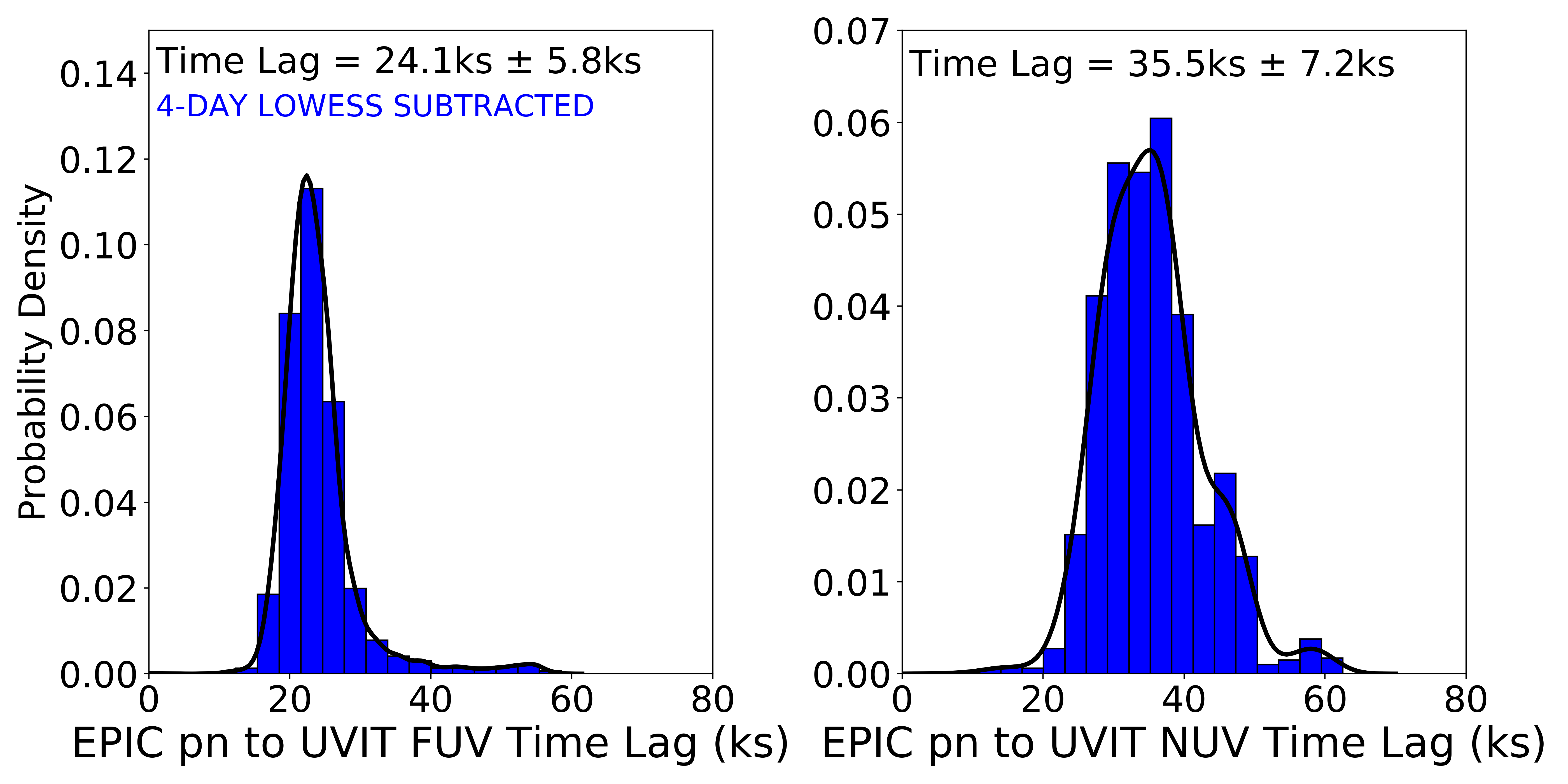}
    \caption{XMM EPIC pn 0.5-10keV to \astrosat Lowess-Detrended UVIT FUV BaF2 (left) and EPIC pn 0.5-10keV to UVIT NUVB4 (right) Cross Correlation Time Lags.}
    \label{fig:XMMtoAstroSat_lags}
\end{figure}

We also measure the lags of the Lowess-Detrended FUV and the NUV relative to the high-intensity XMM EPIC pn X-rays. These lags produce the distributions found in Fig.~\ref{fig:XMMtoAstroSat_lags}, reporting lags of 24.1 $\pm$ 5.8 ks and 35.5 $\pm$ 7.2 ks, for the Lowess-Detrended FUV and the NUV respectively. Both of these lags are consistent with their SXT counterparts within the uncertainties and considering the differences in sampling patterns.

\begin{figure}
    \includegraphics[width=\columnwidth]{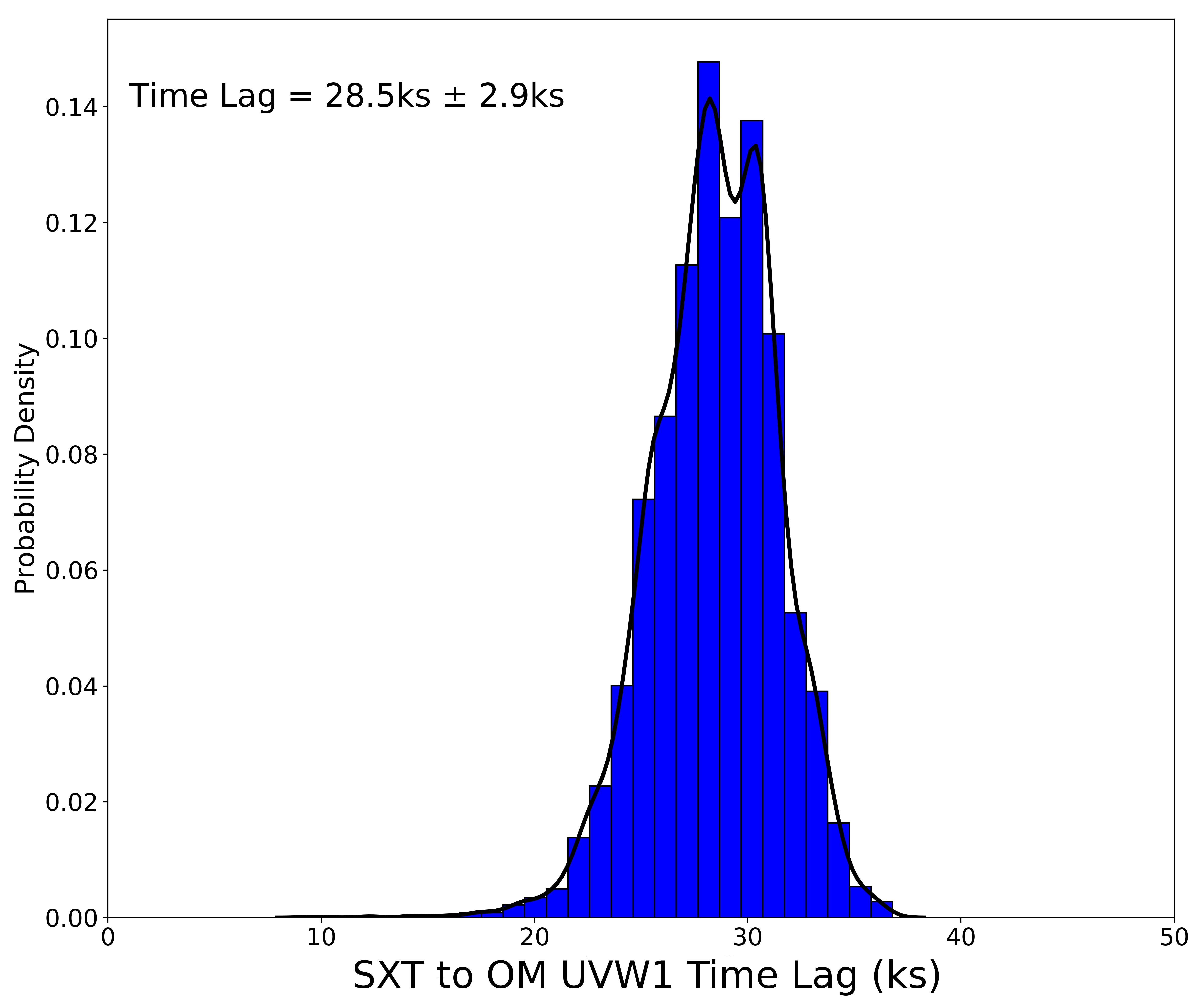}
    \caption{\astrosat SXT 0.3-7keV to XMM OM UVW1 Cross Correlation Time Lag.}
    \label{fig:AstroSattoOM_lag}
\end{figure}

Finally, we can calculate the lag between the \astrosat SXT and the XMM OM UVW1 light curve. The lag distribution is shown in Fig.~\ref{fig:AstroSattoOM_lag}. The measured lag is 28.5 $\pm$ 2.9 ks which agrees with all of the previous lags derived from short datasets, adding further confirmation to the conclusion that, if measured from short lightcurves or with long timescale trends removed, the lag between the X-rays and UVW1 wavelengths is around 20-30 ks. \par

\subsection{Model Lag Spectra} \label{models}

\begin{figure}
    \includegraphics[width=\columnwidth]{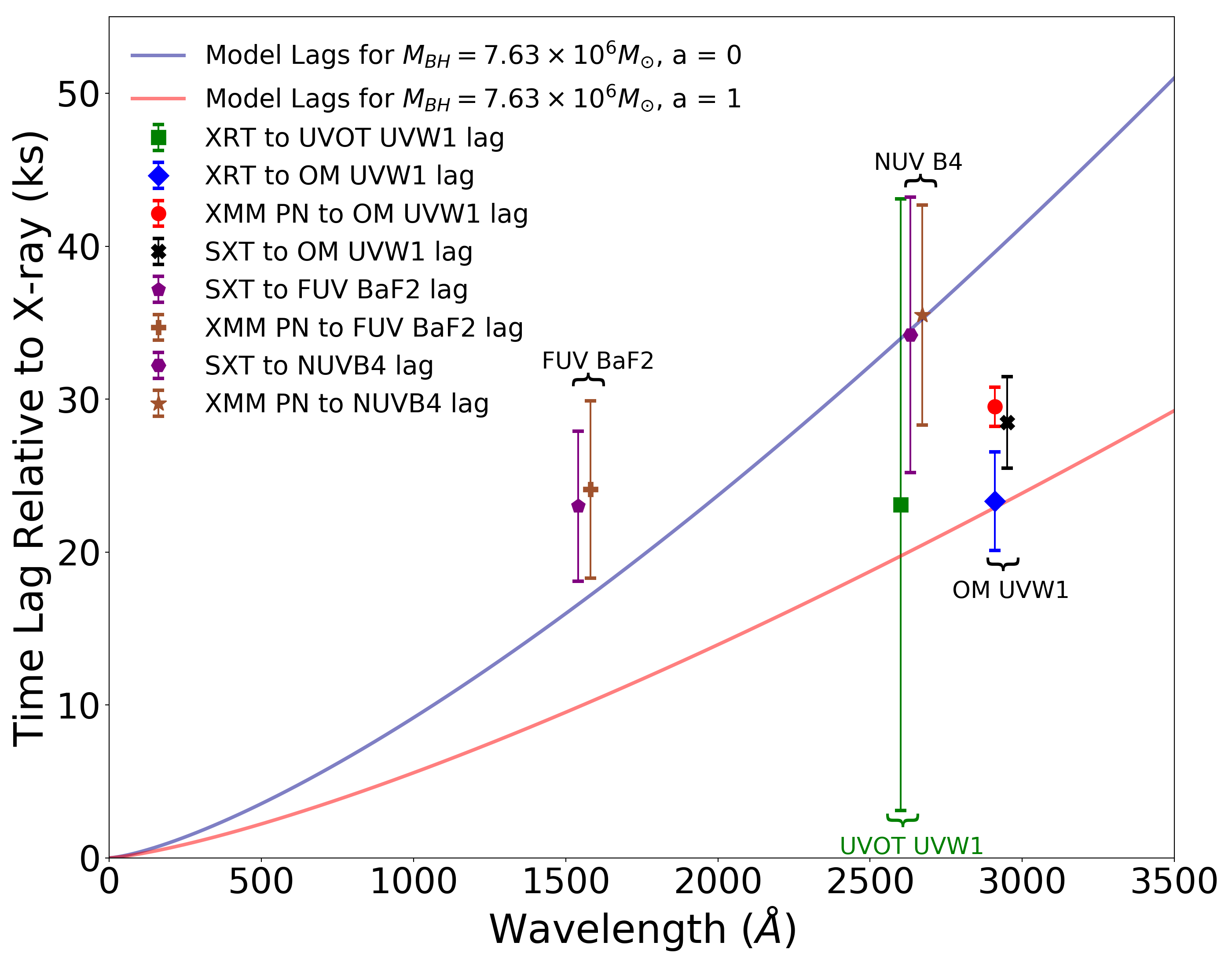}
    \caption{Our UVW1 and UVIT CCF lags compared to the model lag spectra. The XMM to \astrosat UVIT and SXT to OM data points have been shifted 40\si{\angstrom} to the right for visibility.}
    \label{fig:Modelling}
\end{figure}

To test whether the lags measured here are consistent with reprocessing of X-rays directly by a surrounding accretion disc we compare the lag measurements to model predictions. Here we use the analytic model approximation given by \cite{kammoun_2021}, which is derived from the simulation code  KYNreverb, developed by Dov\v{c}iak\footnote{Available from \url{https://projects.asu.cas.cz/stronggravity/kynreverb}}. We assume a coronal lamp-post height of 6 gravitational radii, consistent with that found from X-ray reverberation mapping \citep{emmanoulopoulos_2014, cackett_2014} and microlensing \citep{dai_2009,mosquera_2013}, a mass of  $7.63\times10^{6}$ M$_{\odot}$ and an accretion rate in Eddington units of 8.1\% as listed in Table 3 of \cite{mchardy_et_al_2018}. From the same source we also derive an illuminating 2-10keV luminosity of $6.31\times10^{42}$ ergs s$^{-1}$, taken from the 0.1-195keV ionising BAT luminosity of $3.0\times10^{43}$ using a correction factor of 0.288 assuming a spectrum with an index of $\Gamma = 2$. The predicted lags for both minimum (a=0) and maximum (a=0.998, but approximated in the model by a=1) spin, together with the observed lags, are shown in Fig.~\ref{fig:Modelling}.\par 

The lags in Fig.~\ref{fig:Modelling} are not completely independent as the same UV observations are used in conjunction with different X-ray observations. We indicate the relevant UV observations on the figure. The lags using the \swift UVW1 (though with very large error) and the \xmm OM observations favour the high spin option but the \astrosat NUVB4 observations favour low spin. An intermediate spin would fit these data better, with the exception of FUV BaF2, whose lag appears too large but will be explored in our full lag spectrum. We do not draw strong conclusions from these data except to note that, whatever the spin, the previously accepted mass and accretion rate are mostly consistent with these lags. We also note that a lag fit that goes straight to the origin is quite consistent with the data.

\begin{figure}
    \includegraphics[width=\columnwidth]{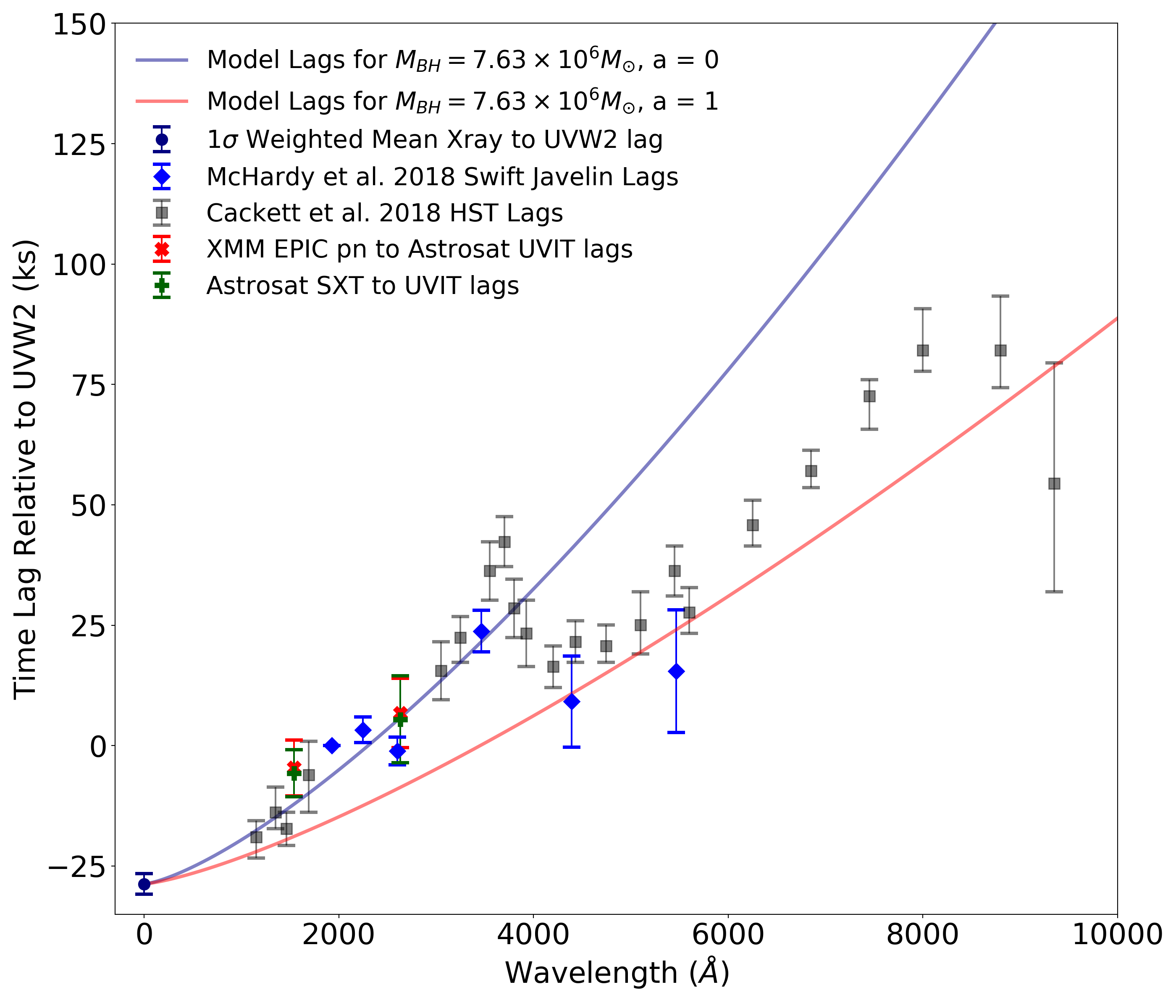}
    \caption{\swift lags for NGC 4593 relative to \protect\cite{mchardy_et_al_2018} UVW2 lags, as well as HST lags from \protect\cite{cackett_2018} and our \astrosat lags.}
    \label{fig:NewLagSpectrum}
\end{figure}

To investigate the lag modelling further we add the UV inter-band lags from \cite{mchardy_et_al_2018} and \cite{cackett_2018}, referenced to our new X-ray to UVW1 lag, taking the weighted mean of all of those lags. It should be noted that although these previous lags were not taken from smoothed light curves, we have not seen evidence of bimodality in inter-UV lags, and so as the previous \swift and HST lags used here were all measured relative to the UVW2 band, the lack of smoothing should not interfere. The resultant combination is shown in Fig.~\ref{fig:NewLagSpectrum}, with all lags plotted relative to UVW2.

By referencing the \swift UVOT and HST lags to our new best estimate of the lag of the X-rays by the UVW1, we see more clearly that the whole lag spectrum does continue smoothly to the X-ray origin without need for an additional offset such as has previously been commonly found. We also see that, assuming the previously accepted mass of $7.63\times10^{6}$ M$_{\odot}$ and an accretion rate in Eddington units of 8.1\%, a low spin model could fit the lags out to $\sim$3500\si{\angstrom}. However such a model overpredicts the lags at longer wavelengths. A high spin model slightly underpredicts the majority of the lags, except around 5000\si{\angstrom} where there are fewer emission lines. There are, however, clear excesses in the 2000-4000\si{\angstrom} and 7000-8000\si{\angstrom} region. These excesses have been noted previously as probably being due to Balmer and Paschen continuum emission, respectively \citep[e.g.][]{cackett_2018} (see Appendix \ref{appendix}). This emission is believed to originate from reprocessing of high energy emission in the Broad Line Region \citep{korista_goad_2001,korista_goad_2019}. This also helps explain the excess lag in the FUV measurements, as by comparing it to the lags around it from \cite{cackett_2018} it appears this BLR component is contributing to lag excess at this waveband. Thus, overall, a high spin model with additional BLR contribution provides the best explanations of the observed lags.
\par

\section{Conclusions} \label{conclusion}

In this paper we presented a 140ks observation of NGC 4593 using \textit{XMM-Newton}'s EPIC PN and OM cameras which sample the X-ray and 2900\si{\angstrom} UV variability, on short timescales ($\sim$1.6d), much better than any previous data. Hence we measure the lag between these bands, on short timescales, much more accurately than with previous data. The lag which we measure, $29.5 \pm 1.3$ks, is only about half the value of 65 ks measured previously from \swift observations covering a longer observing period (22d) but with worse sampling \citep{mchardy_et_al_2018}. Although we measure the lag using a standard Flux Randomisation Cross Correlation Function, we also show that if the \xmm PN and OM lightcurves are superposed, with the OM shifted back by 29~ks, the two lightcurves overlap extremely well.

We have also reexamined the \swift observations, part of which were simultaneous with the XMM observations. The lag distribution between the \swift X-ray and UVW1 bands is actually bimodal with peaks of 27.0 $\pm$ 1.5 ks and 67.0 $\pm$ 1.5 ks. The \swift UVW1 central wavelength (2600\si{\angstrom}) is slightly different to that of the XMM OM (2910\si{\angstrom}), and so lags to the \swift UVW1 might be expected to be $\sim$13$\%$ smaller than for the XMM OM UVW1, assuming lag $\propto \lambda^{4/3}$. The first peak is therefore consistent with our XMM lag and the second is consistent with the longer lag listed by \cite{mchardy_et_al_2018}. 

In addition we analysed 4d of observations with \astrosat's SXT (X-ray) and UVIT (far and near UV) cameras which again overlapped with both the XMM and \swift observations.  The central wavelength of the NUVB4 filter, 2632\si{\angstrom}, is almost identical to that of the \swift UVW1. The \astrosat sampling is not as intense as that of XMM and so uncertainties are larger. However the lag of $34.2 \pm 9.0$ks between the \astrosat X-ray and NUVB4 lightcurves is consistent with the XMM OM lag and with the shorter lag measured with \swift UVW1 data. 

We therefore suggest that the two peaks seen in the \swift lag distribution are in fact two components which dominate over different time scales. The two peaks are also found if the \swift data are split into hard and soft spectral components so they are not a result of an energy dependent mechanism. They are also found in both the high and low sampling halves of the \swift observation so these are not lags which vary with the epoch of observation. To further investigate the timescale dependence we subtracted a 5d smoothed light curve from the \swift data using the LOWESS smoothing technique and measured the lags of both the detrended and the smoothed light curves. The detrended light curves gave a single lag peak of 23.8 $\pm$21.2 ks and the smoothed light curves giving a lag of 99.2 $\pm$17.2 ks. \par

Whether these two lags are two discrete lags or different manifestations of one more complex lag distribution containing multiple timescales, where the lag that dominates depends on the timescales under investigation, is not clear from just these data. Previous observations by \cite{pahari_2020} found two very different X-ray to UV lags in NGC~7469 depending on whether long timescale trends were removed or not. When not removed the UV led the X-rays by $3.5 \pm 0.2$d and when they were removed the UV lagged by $0.37 \pm 0.14$d. The short timescale UV lag was consistent with reprocessing from a disc but a UV lead requires a completely different mechanism, possibly propagating disc fluctuations \citep{arevalo_2006_investigating}. It is not easy to explain these two very different lags as all being part of one broad lag distribution from one broadly similar process. It is simpler to explain them as resulting from two different physical processes. However examination of the \swift observations discussed here using MEMecho analysis showed that response functions with a strong short timescale lag peak but a longer, weaker tail, could explain the observations \citep{mchardy_et_al_2018}. If such response functions did underlie the variability we have observed here, then the short XMM observations would not pick up variations on timescales of a few days. \par

Nonetheless the strong short timescale lag which we detect here is real. If we shift the XMM OM lightcurve by the longer \swift lag, the X-ray and UV lightcurves do not overlap at all well.  The short (29 ks) lag is consistent with reprocessing from an accretion disc directly illuminated from a central X-ray emitting corona of size ($\sim 6Rg$) compatible with measurements from X-ray reverberation and microlensing. Thus we conclude that such direct illumination does occur in at least NGC~4593 and NGC~7469 and produces at least some of the UV and optical variability.  The longer (65 ks) lag is not compatible with direct illumination of the disc but the extra lag might result from scatterings and delays through an inflated inner disc \citep[e.g.][]{gardner_2017_the}. Perhaps there is a structure a little further out in the disc such as a wind which could reprocess the X-rays. The lag is still rather short for reprocessing in the BLR although if the AGN and disc were of low inclination, and the BLR or disc wind had a polar structure, the additional path length from reprocessing to the observer could be very small. Alternatively the longer lag might just be a different manifestation of a broader lag distribution, as sampled by those particular \swift observations. \par

To extend the lags over a wider wavelength range, and particularly at shorter wavelengths, we also measure the lag of the very short wavelength (1541\si{\angstrom}) \astrosat FUV lightcurve relative to the X-rays, using both the \astrosat SXT X-ray lightcurve and the better sampled \xmm PN, discovering similar larger than expected lags for the waveband in the distributions and visual inspection of the light curve suggested that the measured lags of $\sim$33ks did not fit the data. \par 

Therefore we also subtracted a LOWESS light curve smoothed over the entire length of the light curve ($\sim$4d). The smoothing found a feature very similar to what was detected in \swift over the same period, giving confidence that the trend removed was real. This detrended light curve measures a lag of $\sim$23 ks against the X-rays, which appears to fit with both the light curves themselves and the lag spectrum when the Broad Line Region Balmer component is take into account. \par 

An analytic model fit using the formula from \cite{kammoun_2021} for the previously accepted mass of $7.63 \times 10^{6}$ solar and accretion rate of 8.1 per cent Eddington with maximum spin agrees reasonably well with the measurements. This model extends straight to the X-ray zero point without requiring any additional X-ray offset, such as is commonly seen in other AGN.

To the new lags measured here we then added the published lags from \cite{mchardy_et_al_2018} and \cite{cackett_2018}, referencing all lags to zero at the \swift UVW2 wavelength. The resulting total lags spectrum remains consistent with a model fit through the X-ray zero point without any additional X-ray to far UV offset. Thus we conclude that in this AGN the extra offset published in the observations of \cite{mchardy_et_al_2018} and \cite{cackett_2018} may be a consequence either of the domination in the longer \swift observations of a longer lag component. Alternatively it could be a result of a distortion in the measured shorter lag by the presence of an underling longer timescale trend (e.g. see \cite{welsh_1999}) unrelated to X-ray reprocessing. The removal of such trends has been shown before \citep[e.g.][]{mchardy_2014_swift} to remove the additional offsets. Overall we conclude that a high spin model agrees better with the observations than a low spin model although an intermediate spin would probably fit even better. These model fits also require the additional contribution noted in the earlier observations from the BLR in the form of longer Balmer and Paschen continuum lags. \par

These results, which rely heavily on the \xmm OM UVW1 observations, demonstrate the importance of highly sampled UV data. 

\section*{Acknowledgements}

MWJB acknowledges support from STFC in the form of studentship ST/S505705/1.
IMcH and DK also acknowledge support from STFC from grant ST/V001000/1.
This work has used the data from the \astrosat Soft X-ray Telescope (SXT) developed at TIFR, Mumbai, and the SXT POC at TIFR is thanked for verifying and releasing the data via the ISSDC data archive and providing the necessary software tools. This work made use of data supplied by the UK Swift Science Data Centre at the University of Leicester.

\section*{Data Availability}
The \xmm data is available from \url{http://nxsa.esac.esa.int/nxsa-web/#home}. The \swift data is available from \url{https://www.swift.ac.uk/archive/}. The \astrosat data is available from \url{https://astrobrowse.issdc.gov.in/astro_archive/archive/Home.jsp}.




\bibliographystyle{mnras}
\bibliography{ref.bib} 



\appendix

\section{Residual Model Plot} \label{appendix}

\begin{figure}
    \centering
    \includegraphics[width=\columnwidth]{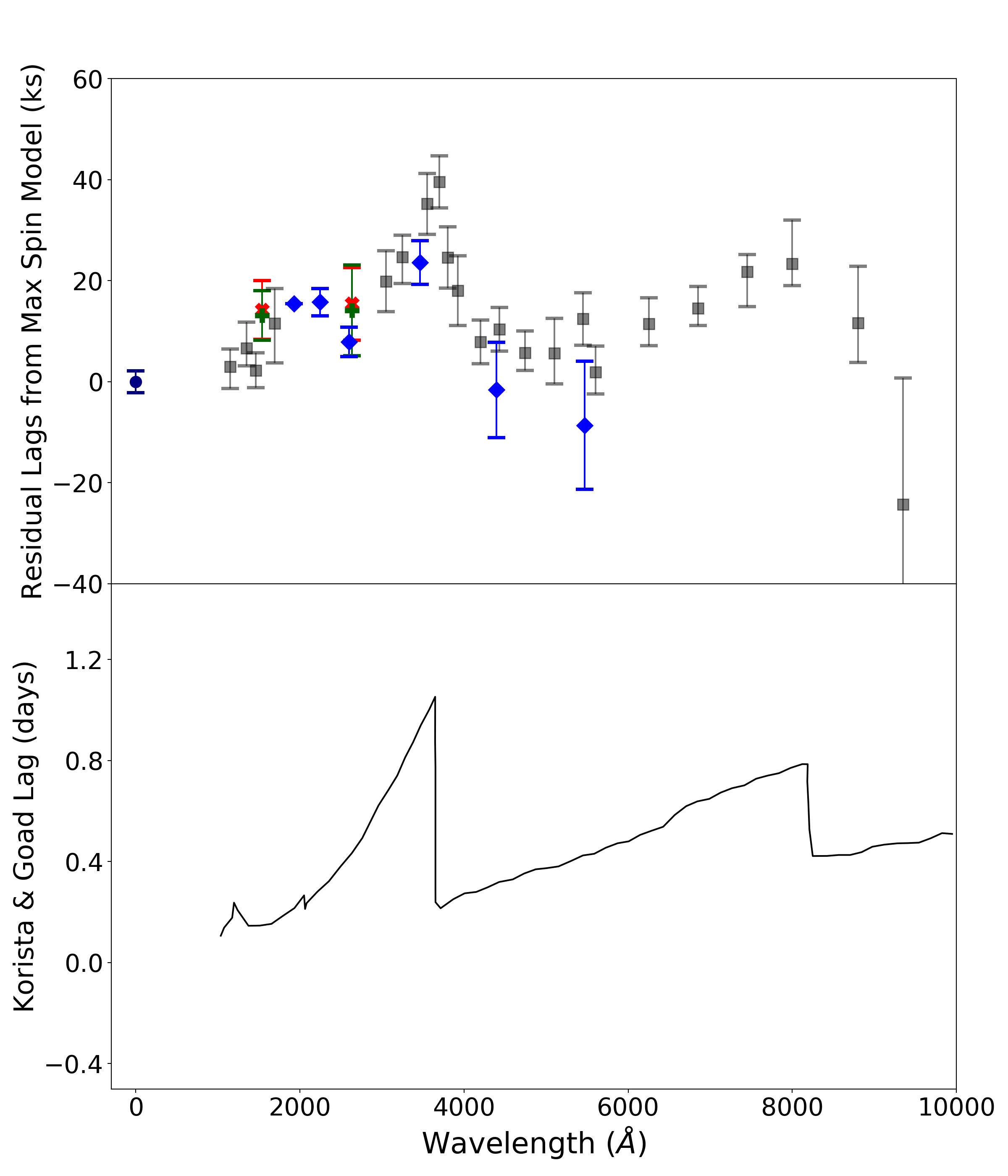}
    \caption{NGC 4593 Residual data lags from the Max Spin Model, and the BLR lag shape from Korista \& Goad 2019.}
    \label{fig:ResidualLagPlot}
\end{figure}

Initially, one might think that the maximum lag model underestimates the lag in the data in the 6000-8000\AA~range as there is an apparent lag excess. This is in fact the expected part of the Paschen component of the BLR lag excesses we observe in this lag spectrum, though it is less visually apparent than the Balmer excess. To demonstrate this we have plotted the residual data lags from the Maximum Spin KYNreverb model and plotted them alongside the expected shape of BLR lag excesses as seen in \cite{korista_goad_2019}, as can be seen in Fig.~\ref{fig:ResidualLagPlot}.

The lag values themselves are different as the \cite{korista_goad_2019} plot is not for NGC 4593, however it is, more importantly, demonstrative of the shape expected from the BLR contribution. While the Balmer excess at $\sim$4000\AA~is much more visually obvious, the Paschen excess should also be present, and so a lower spin model that had the model spectrum intersecting the data points above 6000\AA~would likely be fit incorrectly, as it would be removing the Paschen component of the BLR contribution.


\label{lastpage}
\end{document}